\pdfoutput=1

\documentclass[twoside,leqno,twocolumn]{article}

\usepackage[letterpaper]{geometry}

\usepackage{ltexpprt}

\usepackage[normalem]{ulem}

\newcommand{\boldA}{{\boldsymbol{A}}}

\newcommand{\boldS}{{\boldsymbol{S}}}

\newcommand{\bolda}{{\boldsymbol{a}}}

\newcommand{\boldc}{{\boldsymbol{c}}}

\newcommand{\bolde}{{\boldsymbol{e}}}

\usepackage{bbm}
\usepackage[ruled, algo2e, vlined]{algorithm2e}

\SetCommentSty{mycommfont}

\usepackage[utf8]{inputenc} 
\usepackage[T1]{fontenc}    
\usepackage{hyperref}       
\usepackage{url}            
\usepackage{booktabs}       
\usepackage{amsfonts}       
\usepackage{nicefrac}       
\usepackage{microtype}      
\usepackage{xcolor}         

\hypersetup{colorlinks, citecolor=blue}

\usepackage[ruled,vlined]{algorithm2e}
\usepackage{setspace}

\usepackage{epsfig,bm,subfigure,graphicx,color}
\usepackage{subfigure}
\usepackage{wrapfig}

\usepackage[normalem]{ulem}

\usepackage{amsmath,amssymb,amsfonts,mathtools}
\usepackage{bbm}
\usepackage{lscape}

\usepackage[font=small,skip=0.2cm]{caption}

\begin{document}

\newcommand\relatedversion{}
\renewcommand\relatedversion{\thanks{The full version of the paper can be accessed at \protect\url{https://arxiv.org/abs/2105.12353}}} 

\title{\Large Private Recommender Systems: How Can Users Build Their Own Fair Recommender Systems without Log Data?}
\author{Ryoma Sato\thanks{Kyoto University / RIKEN AIP.}}

\date{}

\maketitle


\fancyfoot[R]{\scriptsize{Copyright \textcopyright\ 2022 by SIAM\\
Unauthorized reproduction of this article is prohibited}}





\begin{abstract} \small\baselineskip=9pt Fairness is a crucial property in recommender systems. Although some online services have adopted fairness aware systems recently, many other services have not adopted them yet. In this work, we propose methods to enable the users to build their own fair recommender systems. Our methods can generate fair recommendations even when the service does not (or cannot) provide fair recommender systems. The key challenge is that a user does not have access to the log data of other users or the latent representations of items. This restriction prohibits us from adopting existing methods designed for service providers. The main idea is that a user has access to unfair recommendations shown by the service provider. Our methods leverage the outputs of an unfair recommender system to construct a new fair recommender system. We empirically validate that our proposed method improves fairness substantially without harming much performance of the original unfair system.\end{abstract}

\section{Introduction}

Fair recommender systems have attracted much attention owing to their importance in society \cite{milano2020recommender}. A typical application is in the job market \cite{geyik2019fairness, feldman2015certifying, geyik2018building}.
The disparate impact theory prohibits a recruiting process that has an adverse impact on a protected group, even if the process appears neutral on its face.
Therefore, job recruiters must avoid using unfair talent recommender systems to remove (possibly unintended) biases in their recruiting process.

However, even if users of a service want to adopt fair recommender systems, they cannot utilize them if the service does not provide them. There are several difficulties facing the adoption of fair recommender systems. First, commercial services may be reluctant to implement fair systems because fairness and effectiveness are in a trade-off relation \cite{mehrotra2018towards, corbett2017algorithmic}, and fairness-aware systems are expensive for implementation and maintenance. Although some social networking services, such as LinkedIn \cite{geyik2019fairness, geyik2018building}, provide fair account recommendations, those of other services, such as Twitter and Facebook, are not necessarily fair with respect to gender or race.
Second, the fairness criterion required by a user differs from user to user. The fairness defined by the service may not match the criteria users call for. For example, even if the recommender system is fair with respect to gender, some users may require fairness with respect to race or the combination of gender and race. In general, different fairness criteria are required in different societies.
Third, companies may refuse to disclose the algorithms they use. This makes it difficult for users to assess the fairness of the system. Milano et al. \cite{milano2020recommender} pointed out that ``The details of RS currently in operation are treated as highly guarded industrial secrets. This makes it difficult for independent researchers to access information about their internal operations, and hence provide any evidence-based assessment.'' In summary, it is difficult for users eager to enjoy fair systems to ensure fairness if they use a recommender system provided by a service.

In this paper, we propose a framework where each user builds their own fair recommender system by themselves. Such a system can provide recommendations in a fair manner each user calls for. In this framework, a user uses their own recommender system via a browser add-on instead of the recommender system provided by the service. We call this personal recommender system a \emph{private recommender system}.

In this work, we focus on item-to-item collaborative filtering, where a set of related items are recommended when a user visits an item page. Each item has a discrete sensitive attribute, such as gender and race. A recommender system must treat all sensitive groups equally.
Examples of this setting include:
\begin{itemize}
    \item \textbf{Job recruiting.} Here, a user is a recruiter, an item is a job seeker, and each job seeker has a sensitive attribute, such as gender and race.
    \item \textbf{Breaking the filter bubble \cite{pariser2011filter}.} Some recommender systems filter information too aggressively. For example, a news recommender system may recommend only conservative news to conservative users \cite{pariser2011filter}. Some users may want to receive unbiased recommendations with respect to ideology. In this case, the user is a reader, the item is a news article, and each news has an ideology label as the sensitive attribute.
    \item \textbf{Avoiding popular item bias.} Recommender systems tend to recommend popular items too much \cite{xiao2019beyond, mehrotra2018towards}. Some knowledgeable users need not receive ordinary items, and they may want to receive uncommon items. For example, IMDb recommends Forrest Gump, The Dark Knight, The Godfather, Inception, Pulp Fiction, and Fight Club in the Shawshank Redemption page\footnote{\url{https://www.imdb.com/title/tt0111161/}}. However, most cinema fans are already familiar with all of these titles, and these recommendations are not informative to them. In this case, we can set the sensitive attribute to be a popularity label (e.g., high, middle, and low, based on the number of reviews received), such that recommender systems must recommend uncommon (but related) items as well. Although readers may think of this problem as topic diversification \cite{ziegler2005improving}, we discuss these problems in a unified framework.
\end{itemize}
We assume that users have access to the sensitive attribute, but they do not necessarily have access to the content of items \cite{koren2009matrix}. Although there exist several methods for building fair recommender systems \cite{singh2018fairness, patro2020fairrec, liu2019personalized}, all of the existing fair recommendation methods are designed for service providers, who can access the entire log data, such as the rating matrix or browsing history of all users. In our setting, however, a user does not have access to the log data of other users nor latent representations of items. A clear distinction between this work and previous works is that our setting prohibits accessing such log data. This restriction makes the problem challenging. The key idea is that a user has access to unfair recommendations shown by the service provider. We propose methods to leverage the outputs of an unfair recommender system to construct a fair recommender system. We conduct several experiments and show that our proposed method can build much fairer recommender systems than provider recommender systems while keeping their performance. The contribution of this paper is as follows:
\begin{itemize}
    \item We propose private recommender systems, where each user builds their own recommender system to ensure the fairness of recommendations. Private recommender systems enable fair recommendations in many situations where conventional recommendation algorithms cannot be deployed. Our proposed framework expands the application scope of fairness-aware recommender systems.
    \item We propose methods to develop private recommender systems without accessing log data or the contents of items. Although our methods are simple, they exhibit a positive trade-off between fairness and performance, even without accessing log data.
    \item We confirm that our proposed method works in real-world scenarios via qualitative case studies on IMDb and Twitter.
\end{itemize}

\noindent \uline{\textbf{Reproducibility:}} Our code is available at \url{https://github.com/joisino/private-recsys}.

\section{Notations}

For every positive integer $n \in \mathbb{Z}_+$, $[n]$ denotes the set $\{ 1, 2, \dots n \}$.
Let $K \in \mathbb{Z}_+$ be the length of a recommendation list. Let $\mathcal{U} = [m]$ denote the set of users and $\mathcal{I} = [n]$ denote the set of items, where $m$ and $n$ are the numbers of users and items, respectively. Without loss of generality, we assume that the users and items are numbered with $1, \dots, m$ and $1, \dots, n$, respectively.

\section{Problem Setting} \label{sec: setting}

We focus on item-to-item collaborative filtering. In this setting, when user $u$ accesses the page associated with item $i$, a recommender system aims to recommend a set of items that are relevant to item $i$. The recommendation may be personalized for user $u$ or solely determined by the currently displayed item $i$. We assume that the recommendation list does not contain any items that user $u$ has already interacted with. This is natural because already known items are not informative. Formally, a recommendation list is represented by a $K$-tuple of items, and a recommender system is represented by a function $\mathcal{F}\colon \mathcal{U} \times \mathcal{I} \to \mathcal{I}^K$ that takes a user and source item and returns a recommendation list when the user visits the source item. $\mathcal{F}(u, i)_k$ denotes the $k$-th item of $\mathcal{F}(u, i)$.

We assume that each item $i$ has a sensitive attribute $a_i \in \mathcal{A}$, and we can observe the sensitive attribute. $\mathcal{A}$ is a discrete set of sensitive groups. When we cannot directly observe $a_i$, we estimate it from auxiliary information. A recommender list is fair if the proportion of protected attributes in the list is approximately uniform. For example, ({\color[HTML]{FF4B00} man}, {\color[HTML]{FF4B00} man}, {\color[HTML]{FF4B00} man}, {\color[HTML]{FF4B00} man}, {\color[HTML]{005AFF} woman}, {\color[HTML]{FF4B00} man}) is not fair with respect to $\mathcal{A} = \{\text{\color[HTML]{FF4B00} man}, \text{\color[HTML]{005AFF} woman} \}$ because much more men accounts (items) are recommended than women accounts. We found that LinkedIn employs this kind of fairness in their recommendations. Specifically, at least two men and two women users were recommended out of five recommendation slots in a user's profile page as far as we observed. However, we observed that Twitter does not employ this kind of fairness. An example of our goal in this paper is to develop a fair recommender system on Twitter, although they do not provide a fair recommender system.
In the experiments, we measure fairness quantitatively by least ratio (i.e., the minimum ratio of protected attributes in the list) and entropy. For example, least\_ratio({\color[HTML]{FF4B00} man}, {\color[HTML]{FF4B00} man}, {\color[HTML]{FF4B00} man}, {\color[HTML]{FF4B00} man}, {\color[HTML]{005AFF} woman}, {\color[HTML]{FF4B00} man}) $= 1/6$, and the entropy is $-1/6 \log 1/6 - 5/6 \log 5/6 \approx 0.65$. The higher these values are, the fairer the list is considered to be. In the LinkedIn example above, the least ratio was ensured to be at least $2/5 = 0.4$. If the numbers $n_a$ of men and women are the same, \begin{align*}
    &\text{least\_ratio} = 0.5 \\
    &\Longleftrightarrow \frac{K}{2} \text{ men and } \frac{K}{2} \text{ women are recommended} \\
    &\Longleftrightarrow \text{Pr}[Y \mid A = \text{man}] = \frac{K}{2n_a} = \text{Pr}[Y \mid A = \text{woman}] \\
    &\Longleftrightarrow \text{demographic parity}
\end{align*} holds, where $Y$ is a binary outcome variable that indicates whether an item is included in the recomendation list. When the numbers of men and women are different, we can slightly modify the definition and algorithm so that it includes demographic parity, yet we stick to the above definition in this paper for ease of exposition. Note that these fairness scores do not take the order of recommendation lists into account, but only count the number of items. It is straightforward to incorporate order-aware fairness in our proposed methods by adjusting the post processing process (e.g., adopting \cite{zehlike2017fair}). We leave this direction for future work for ease of exposition.

We assume a recommender system $\mathcal{P}_{\text{prov}}$ in operation provided by a service is available. We call this recommender system a \emph{provider recommender system}. The provider recommender system is arbitrary and may utilize the contents of items by accessing exclusive catalog databases, as well as log data of all users. Although the provider recommender system suggests highly related items, it may be unfair to the protected group. Our goal is to enable each user to build their own fair recommender system. We call this new recommender system a private recommender system. Without loss of generality, we assume that the user who is constructing a private recommender system is user $1$. We assume that user $1$ does not have access to the details of the provider recommender system, including the algorithm, latent representations of items, and score function. The only information that user $1$ can obtain is the top-$K$ ranking of items when user $1$ visits each source item, which is the case in many real-world settings. Formally, let $\mathcal{P}_{\text{prov}, 1}\colon \mathcal{I} \to \mathcal{I}^K$ be the partial application of $\mathcal{P}_{\text{prov}}$ by fixing the first argument to be user $1$, i.e., $\mathcal{P}_{\text{prov}, 1}(i) = \mathcal{P}_{\text{prov}}(1, i) ~\forall i \in \mathcal{I}$. We assume that user $1$ can access function $\mathcal{P}_{\text{prov}, 1}$ by using the provider recommender system. The problem we tackle in this paper is formalized as follows:

\vspace{0.05in}
\noindent \uline{\textbf{Private Recommender System Problem.}}\\
\textbf{Given:} Oracle access to the provider recommendations $\mathcal{P}_{\text{prov}, 1}$. Sensitive attribute $a_i$ of each item $i \in \mathcal{I}$.\\
\textbf{Output:} A private recommender system $\mathcal{Q}\colon \mathcal{I} \to \mathcal{I}^K$ that is fair with respect to $\mathcal{A}$.

\vspace{0.05in}
\noindent \uline{\textbf{Difference with posthoc processing.}} Note that this problem setting is different from posthoc fair recommendations because posthoc processing has access to the full ranking of items or scoring function, whereas we only have access to top-$K$ items presented by the service provider. Although it may be possible to apply posthoc processing to the top-$K$ list, it just alters the order of the list. Therefore, simply applying posthoc processing cannot improve the least ratio. For example, if the top-$K$ list contains only one group, posthoc processing does not recommend items in the other group. If $K$ is sufficiently large, reordering the list by a posthoc processing method and truncating the list may improve the least ratio. However, $K$ is typically small, and we focus on the challenging cases, where recommendation lists are short and may solely contain one group.

\section{Method}

The core idea of our proposed methods is regarding two items $(i, j)$ similar if item $j$ is recommended in item $i$'s page. They recommend items based on this similarity measure while maintaining fair recommendations. However, the similarities of only $K$ pairs are defined for each source item in this manner. The recommendation list may solely contain one protected group. In that case, we cannot retrieve similar items in different groups. To address this challenge, we propose to utilize the recommendation network to define a similarity measure.

\subsection{Recommendation Network}

Recommendation networks \cite{cano2006topology, celma2008new, seyerlehner2009limitation, seyerlehner2009browsing} have been utilized to investigate the property of recommender systems, such as the navigation of a recommender system and novelty of recommendations. The node set of a recommendation network is the set of items, i.e., $V = \mathcal{I}$, and a directed edge $(i, j) \in \mathcal{I} \times \mathcal{I}$ indicates that item $j$ is recommended in item $i$'s page, i.e., $E = \{ (i, j) \in \mathcal{I} \times \mathcal{I} \mid \exists k \text{ s.t. } \mathcal{P}_{\text{prov}, 1}(i)_k = j \}$. The key idea of our proposed methods is that two items can be considered to be similar if they are close in the recommender network. We use this to define the similarities between items. To the best of our knowledge, this is the first work to adopt the recommendation network for defining the similarities between items for constructing recommender systems.

\subsection{PrivateRank} \label{sec: PrivateRank}

In this section, we introduce \textsc{PrivateRank}. First, \textsc{PrivateRank} constructs a recommender network by querying the provider recommender system. We adopt a weighted graph to incorporate the information of item rankings. Inspired by the discounted cumulative gain, the weight function is inverse-logarithmically discounted. The adjacency matrix $\boldA$ is defined as:
\begin{align*}
    \boldA_{ij} = \begin{cases}
    \frac{1}{\log(k + 1)} & (\mathcal{P}_{\text{prov}, 1}(i)_k = j) \\
    0 & \text{(otherwise)}
    \end{cases}
\end{align*}
\textsc{PrivateRank} employs the personalized PageRank \cite{jeh2003scaling, page1999pagerank}, also called the random walk with restart, which is a classic yet powerful similarity measures between nodes on a graph. The personalized PageRank $\boldS_i \in \mathbb{R}^{n}$ of node $i$ measures similarities between node $i$ and other nodes. The personalized PageRank assumes a random surfer who follows a link incident to the current node with probability proportional to its weight or jumps to node $i$ with probability $1 - c$. $c > 0$ is a hyperparameter called a damping factor. The damping factor $c$ controls the spread of random walks. When $c$ is large, the surfer rarely jumps back to the source node, and it solely captures global structures. Therefore, a small $c$ is appropriate in capturing local structures around the source node. We empirically demonstrate it in the experiments. The personalized PageRank of node $j$ with respect to node $i$ is defined as the probability that the random surfer will arrive at node $j$. Formally,
\begin{align*}
    \boldS_{i} = c \tilde{\boldA}^\top \boldS_{i} + (1 - c) \bolde^{(i)}
\end{align*}
\noindent where $\tilde{\boldA}$ is the row-wise normalized adjacency matrix, i.e., $\tilde{\boldA}_i = \boldA_i / \sum_j \boldA_{ij}$, and $\bolde^{(i)}$ is the $i$-th standard basis.
We compute the personalized PageRank using the cumulative power iteration \cite{yoon2018tpa}:
\begin{align*}
    \hat{\boldS}_{i} = (1 - c) \sum_{k = 0}^L (c \tilde{\boldA}^\top)^k \bolde^{(i)},
\end{align*}
where $L$ is a hyperparameter. $L$ determines the trade-off between time consumption and accuracy. We find that a small $L$ is sufficient because recommendation networks are typically small-world. We use $L = 10$ in the experiments.

After we obtain the similarity matrix, \textsc{PrivateRank} ranks items in a fair manner. There are several existing fairness-aware ranking methods, including the optimization-based \cite{singh2018fairness}, learning-based \cite{beutel2019fairness}, and post processing-based approaches \cite{geyik2019fairness, liu2019personalized, zehlike2017fair}. We employ the post processing-based approach similar to \cite{geyik2019fairness, liu2019personalized}. We set the minimum number $\tau ~(0 \le \tau \le K/|\mathcal{A}|)$ of items of each group as a hyperparameter. \textsc{PrivateRank} greedily takes items as long as the constraint can be satisfied. Specifically, let $r$ be the number of items to be taken, and $c_a$ be the number of items in the list with protected attribute $a$. Then, if $\sum_{a \in \mathcal{A}} \max(0, \tau - c_a) \le r$ holds, we can satisfy the minimum requirement by completing the list. The pseudo code of \textsc{PrivateRank} is available in Section \ref{sec: PrivateRankcode} in the appendices. 

\textsc{PrivateRank} holds the following preferable properties. First, it ensures fairness if we increase the minimum requirement $\tau$.

\begin{theorem} \label{thm1}
If $\mathcal{I}$ contains at least $\tau$ items of each sensitive attribute, the least ratio of recommendation list generated by \textsc{PrivateRank} is at least $\tau/K$.
\end{theorem}

The proof is available in Section \ref{sec: thm1} in the appendices.

Second, \textsc{PrivateRank} does not lose performance when $\tau = 0$.

\begin{theorem}
If we set $c < \frac{1}{(K + 1)^2 \log^2(K + 1)}$, $L \ge 1$, and $\tau = 0$, the recommendation list generated by \textsc{PrivateRank} is the same as that of the provider recommender system. Therefore, the recall and nDCG of \textsc{PrivateRank} are the same as those of the provider recommender system.
\end{theorem}

The proof is available in Section \ref{sec: thm2} in the appendices.

This result is consistent with the intuition that small $c$ is good for \textsc{PrivateRank}. These theorems indicate that \textsc{PrivateRank} properly enhances the functionality of the provider recommender system. Specifically, it can recover the original system when $\tau = 0$, and in addition to that, it can control fairness by increasing $\tau$.

\vspace{0.1in}
\noindent \uline{\textbf{Time complexity:}} We analyze the time complexity of \textsc{PrivateRank}. Constructing the recommendation network issues $K$ queries to $\mathcal{P}_{\text{prov}, 1}$ for each item. Therefore, $Kn$ queries are issued in total. Computing an estimate $\hat{\boldS}_i$ of the personalized PageRank involves $L$ vector-matrix multiplications. A vector-matrix multiplication can be done in $O(K n)$ time because $Kn$ elements of $\tilde{\boldA}$ are non-zero. Therefore, computing $\hat{\boldS}_i$ runs in $O(KLn)$ time. The postprocessing runs in $O(K + |\mathcal{A}|)$ time. Hence, constructing the recommendation list takes $O(n (K + |\mathcal{A}|))$ time. In total, \textsc{PrivateRank} runs in $O(n (KL + |\mathcal{A}|))$ time if we assume that evaluating $\mathcal{P}_{\text{prov}, 1}$ runs in a constant time.

\subsection{PrivateWalk}

Although \textsc{PrivateRank} performs well in practice, the main limitation of this method is its scalability. Even if we use a faster approximation method for computing the personalized PageRank, constructing the recommendation network may be a bottleneck of the computation, which requires $Kn$ evaluations of $\mathcal{P}_{\text{prov}, 1}$. Many evaluations of $\mathcal{P}_{\text{prov}, 1}$ may hit the limitation of API, consume too much wall clock time for inserting appropriate intervals, or may be certified as a DOS attack. Therefore, batch methods that construct a full recommendation network are not suitable when many items are involved. We propose an algorithm to build a private recommender system that computes a recommendation list on demand when a user accesses an item page.

The central idea of \textsc{PrivateWalk} is common with that of \textsc{PrivateRank}: two items are similar if they are close in the recommendation networks of the provider recommender system. \textsc{PrivateWalk} utilizes random walks. Two items $(i, j)$ are considered similar if item $j$ can be reached from $i$ in short steps by a random walk. In contrast to \textsc{PrivateRank}, \textsc{PrivateWalk} runs random walks like a web crawler when a user visits a page, rather than building the recommendation networks beforehand. To achieve fairness, \textsc{PrivateWalk} also employs the post processing approach with the minimum requirement $\tau$. The pseudo code of \textsc{PrivateWalk} is available in Section \ref{sec: PrivateWalkcode} in the appendices.

\vspace{0.1in}
\noindent \uline{\textbf{Time complexity:}} The time complexity of \textsc{PrivateWalk} depends on the average length $L_{\text{ave}}$ of random walks, which is bounded by $L_{\text{max}}$. \textsc{PrivateWalk} runs the inner loop (Lines 7--14 in the pseudo code) $K L_{\text{ave}}$ time. In each loop, $\mathcal{P}_{\text{prov}, 1}$ is evaluated once, \texttt{CanAdd} is evaluated once, and $O(K)$ basic operations run. The number of loops of the fallback process (Lines 15--19) is a constant in expectation because the probability decays exponentially with respect to the number of iterations. In total, \textsc{PrivateWalk} runs in $O(K (K + |\mathcal{A}|) L_{\text{ave}})$ time if we assume evaluating $\mathcal{P}_{\text{prov}, 1}$ runs in constant time. The complexity is independent of the number $n$ of items in contrast to \textsc{PrivateRank}.

\section{Experiments}

We will answer the following questions via experiments.

\begin{itemize}
    \item (RQ1) How good trade-off between fairness and performance do our proposed methods strike?
    \item (RQ2) How sensitive are our proposed methods with respect to hyperparameters?
    \item (RQ3) Do our methods work in real-world scenarios?
\end{itemize}

\subsection{Experimental settings}

\begin{figure*}[tb]
\begin{center}
\begin{minipage}{0.20\hsize}
\includegraphics[width=\hsize]{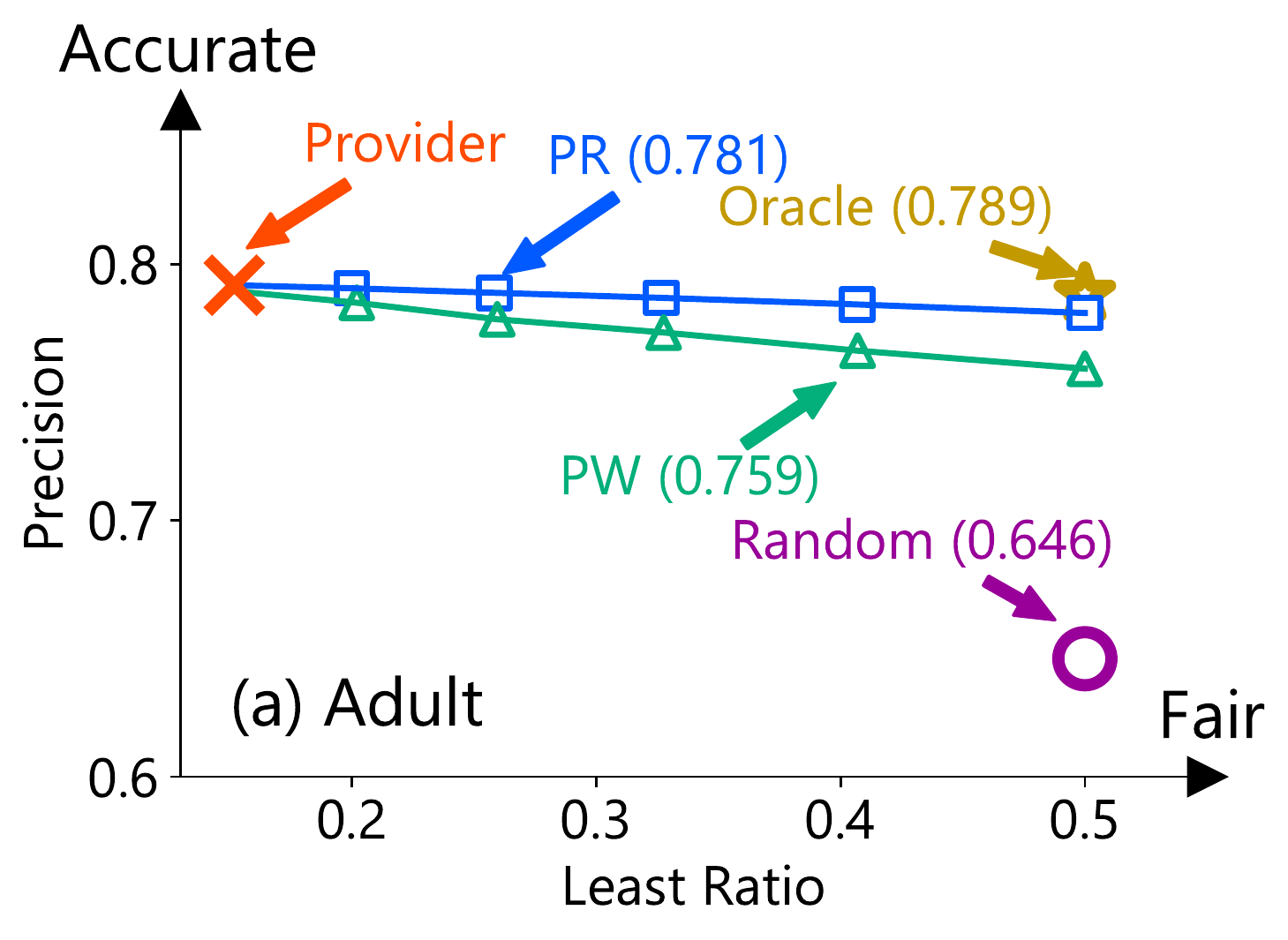}
\end{minipage}
\quad
\begin{minipage}{0.20\hsize}
\includegraphics[width=\hsize]{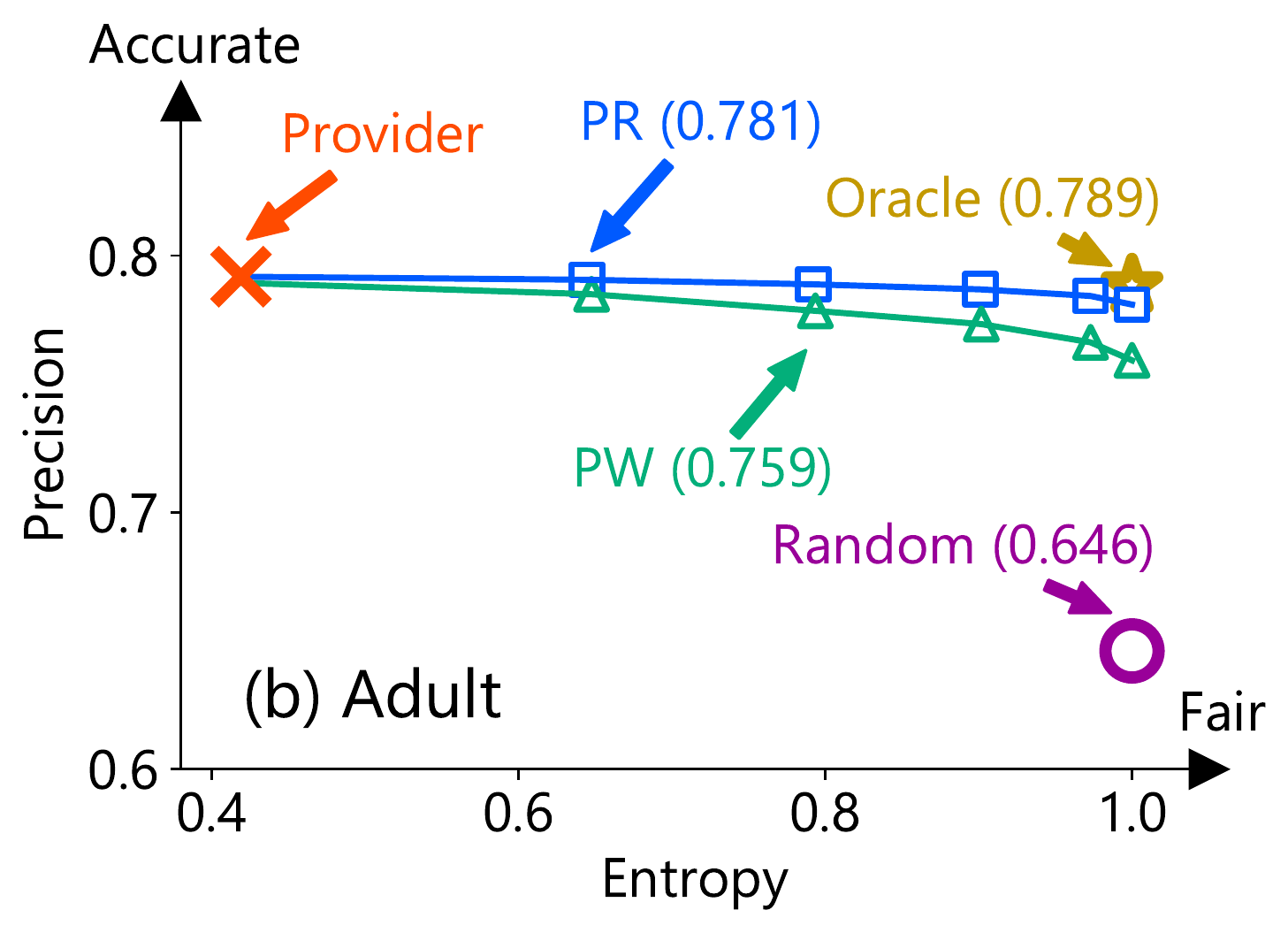}
\end{minipage}
\quad
\begin{minipage}{0.20\hsize}
\includegraphics[width=\hsize]{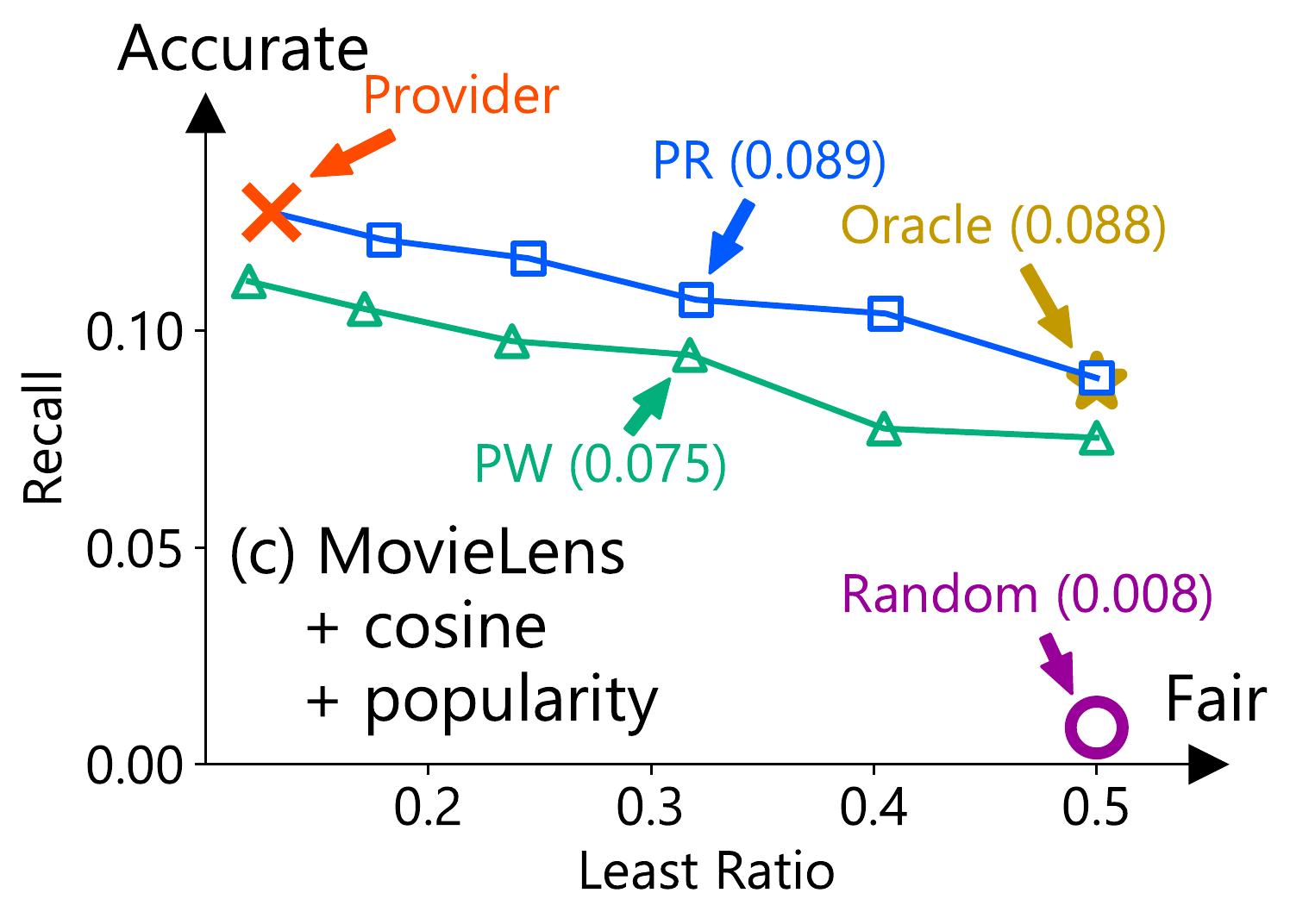}
\end{minipage}
\quad
\begin{minipage}{0.20\hsize}
\includegraphics[width=\hsize]{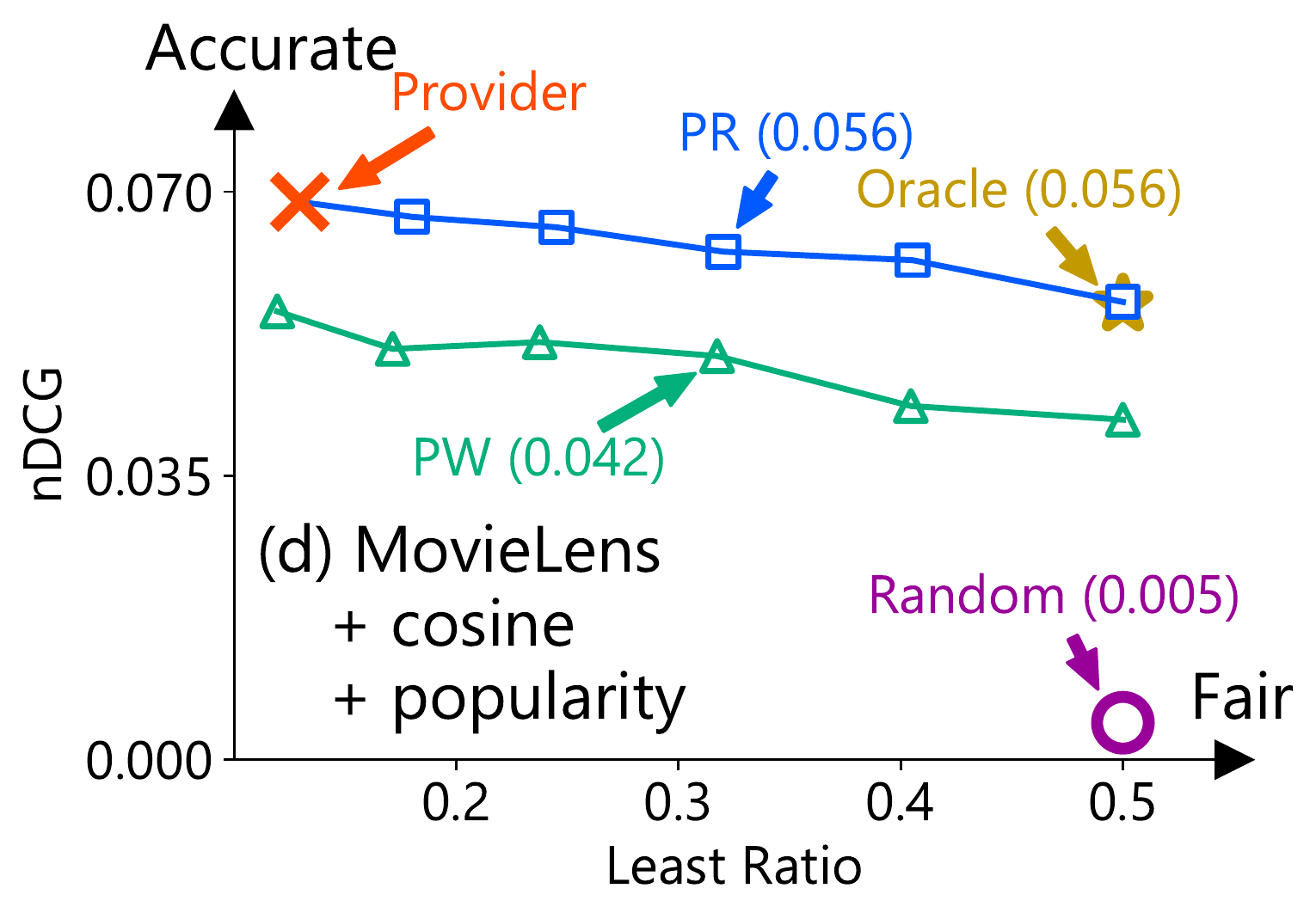}
\end{minipage}
\begin{minipage}{0.20\hsize}
\includegraphics[width=\hsize]{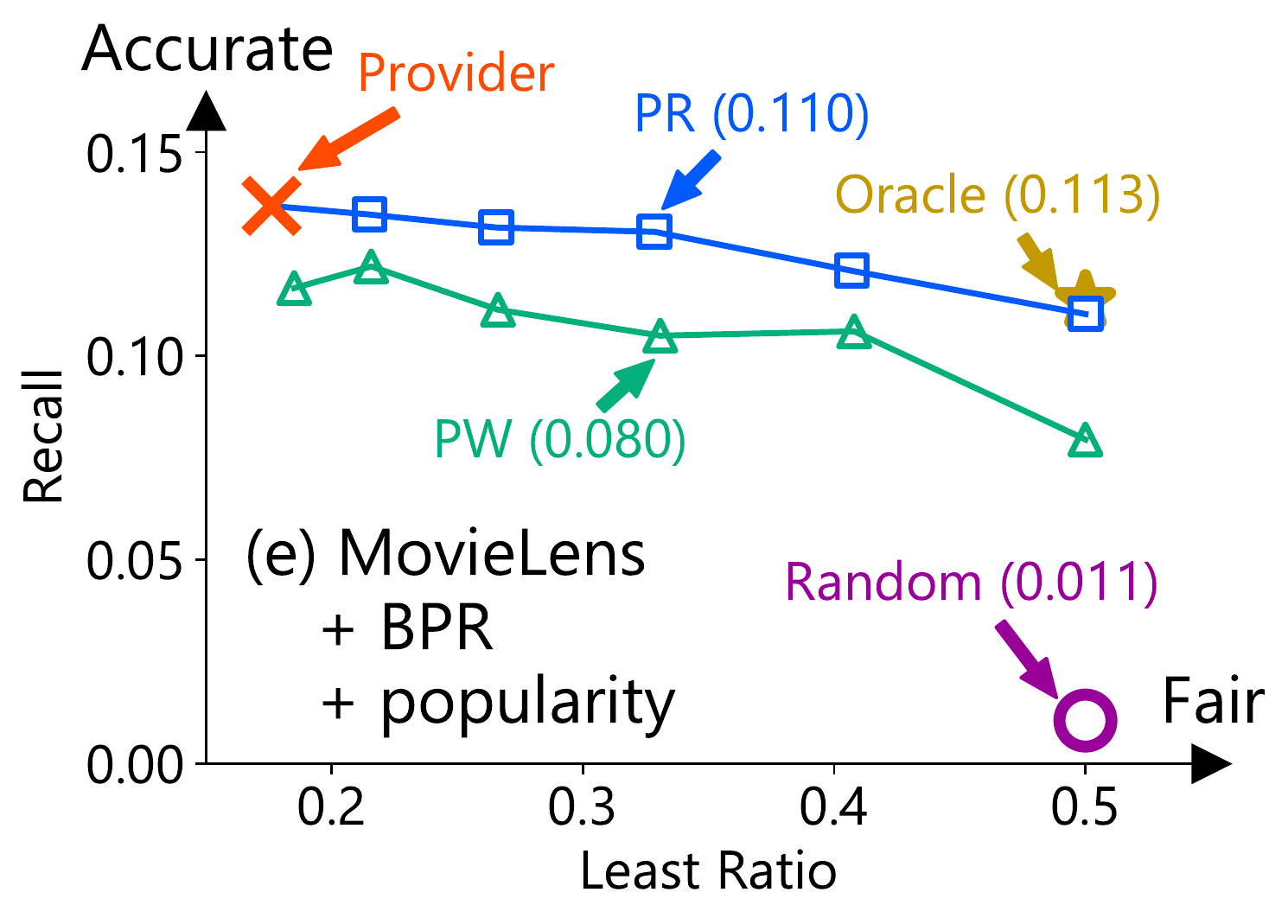}
\end{minipage}
\quad
\begin{minipage}{0.20\hsize}
\includegraphics[width=\hsize]{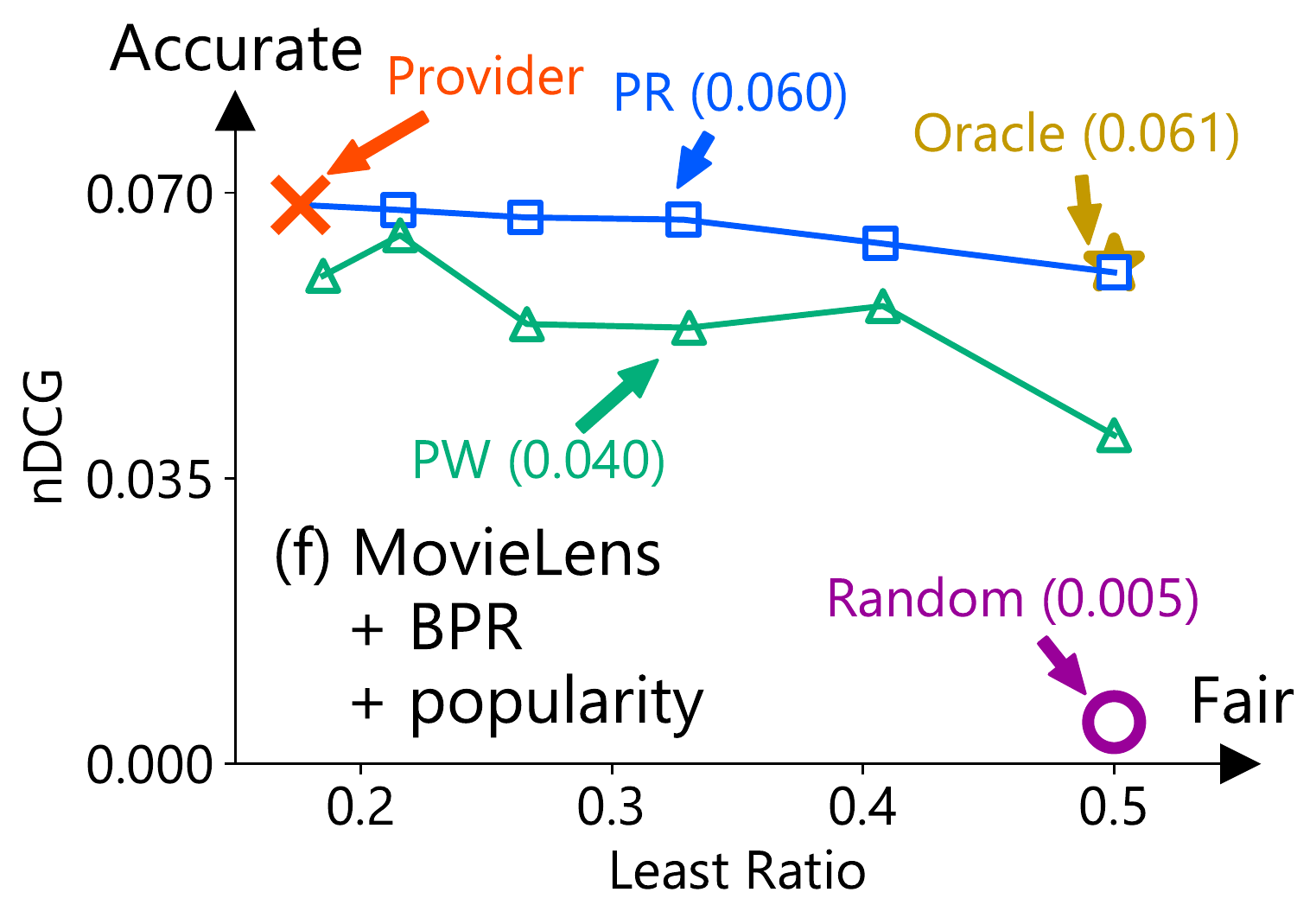}
\end{minipage}
\quad
\begin{minipage}{0.20\hsize}
\includegraphics[width=\hsize]{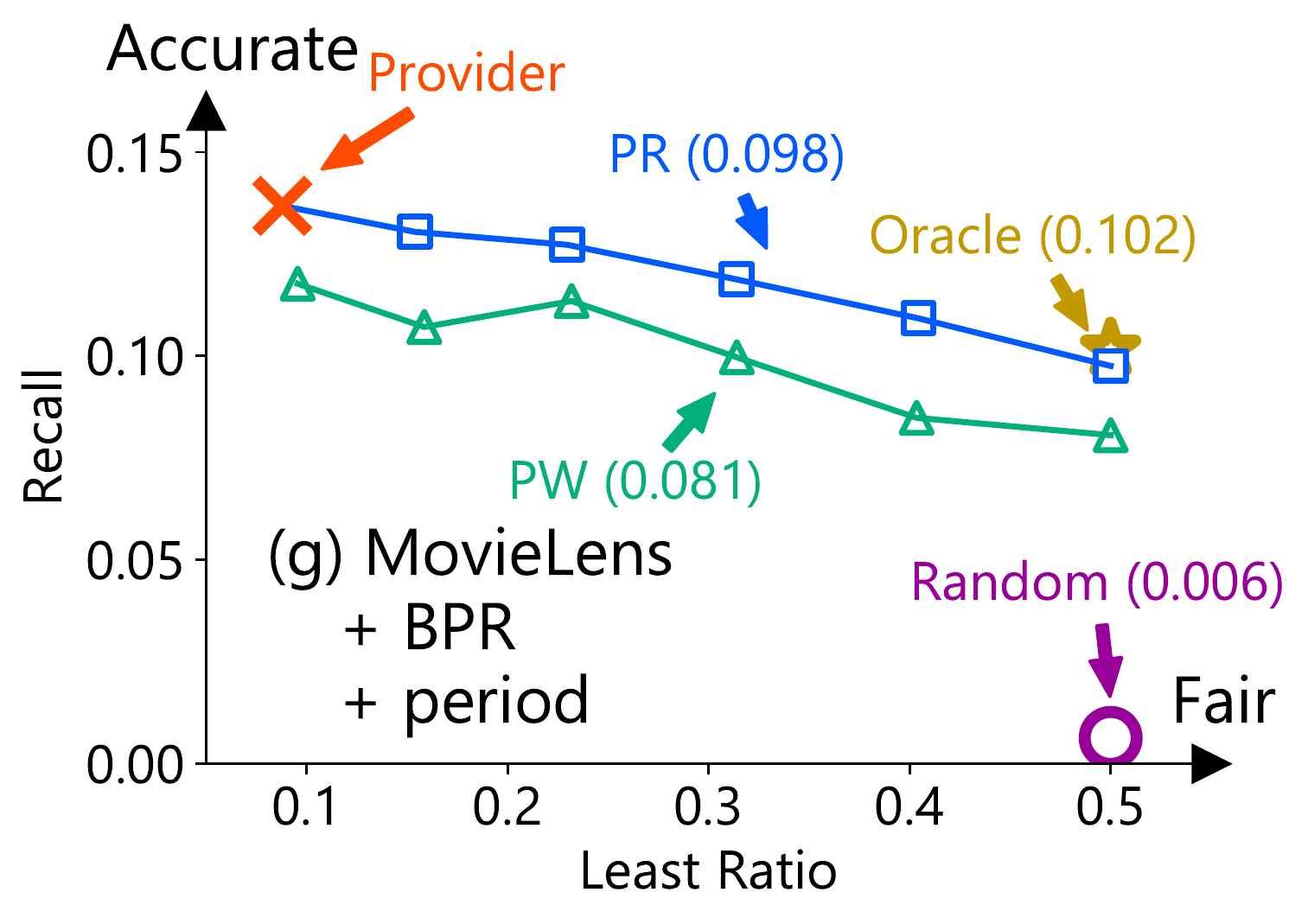}
\end{minipage}
\quad
\begin{minipage}{0.20\hsize}
\includegraphics[width=\hsize]{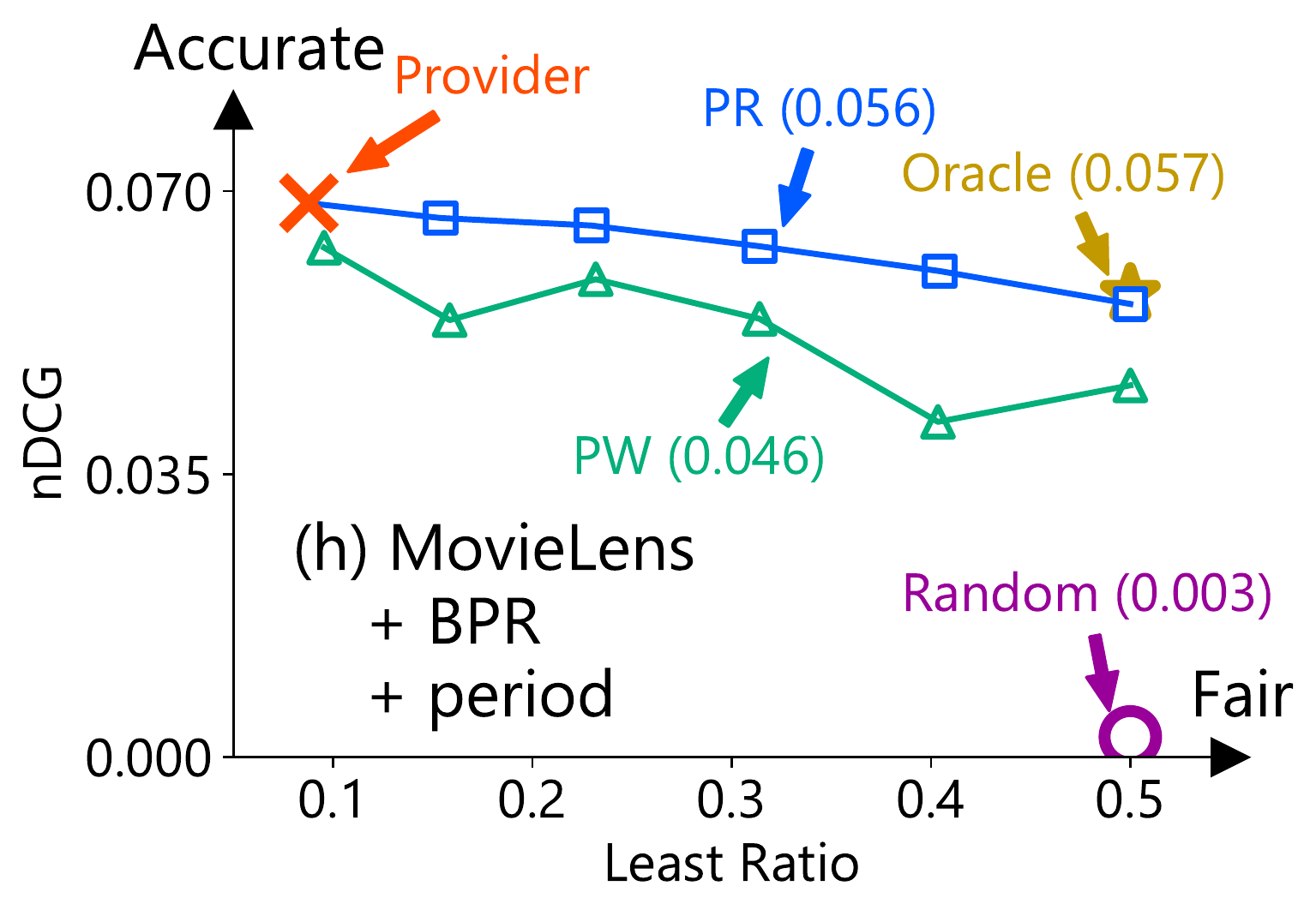}
\end{minipage}
\begin{minipage}{0.20\hsize}
\includegraphics[width=\hsize]{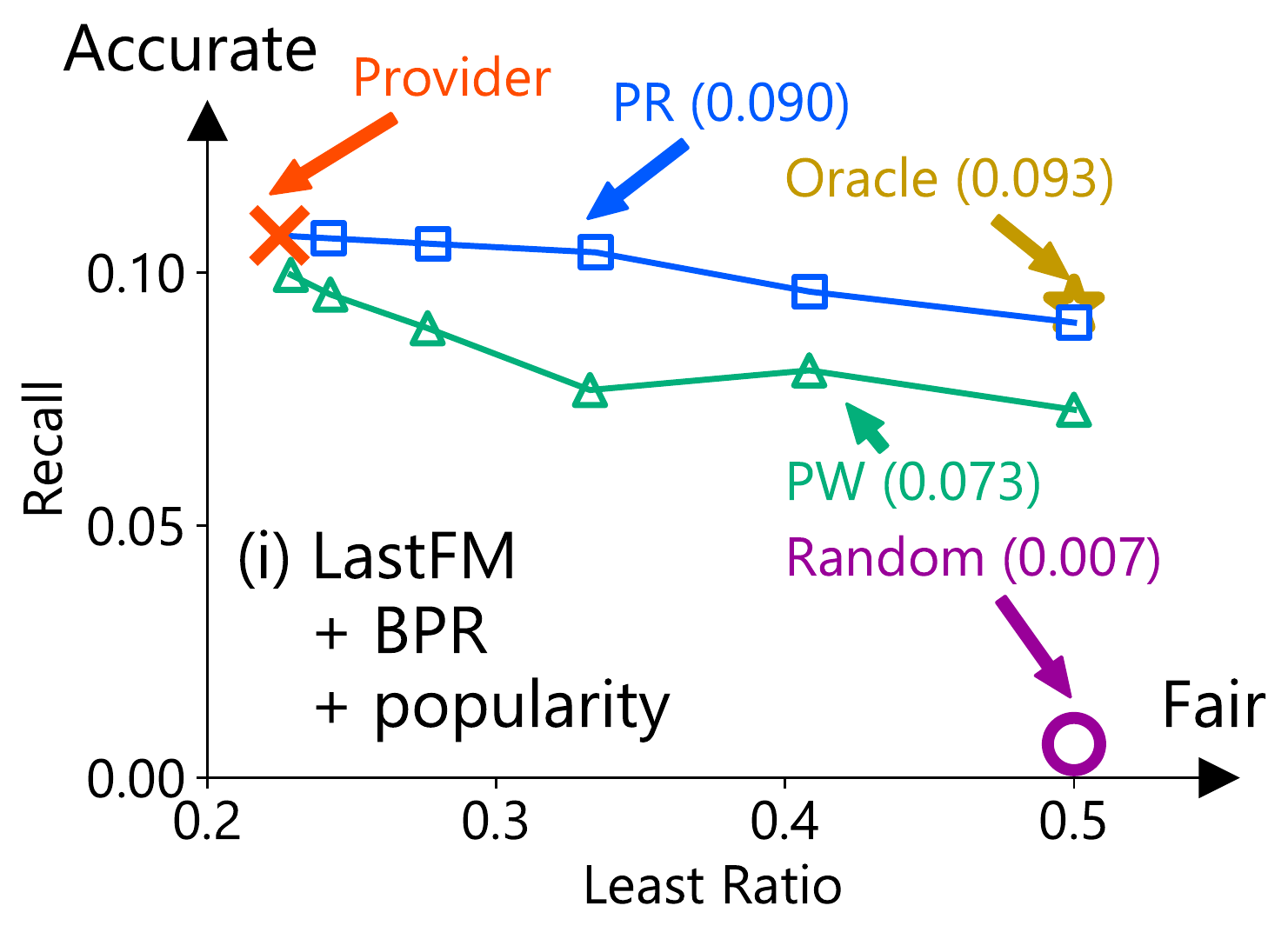}
\end{minipage}
\quad
\begin{minipage}{0.20\hsize}
\includegraphics[width=\hsize]{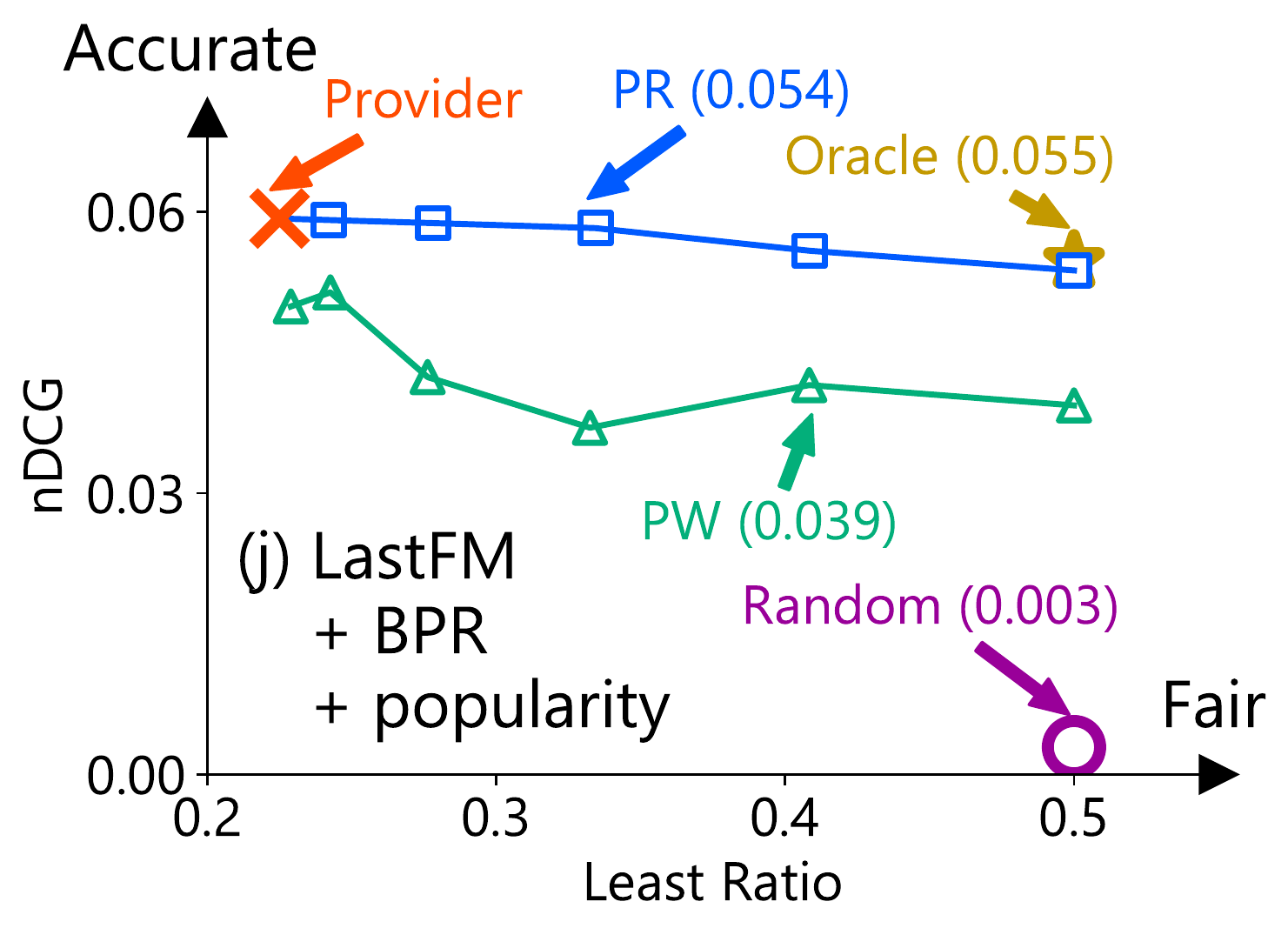}
\end{minipage}
\quad
\begin{minipage}{0.20\hsize}
\includegraphics[width=\hsize]{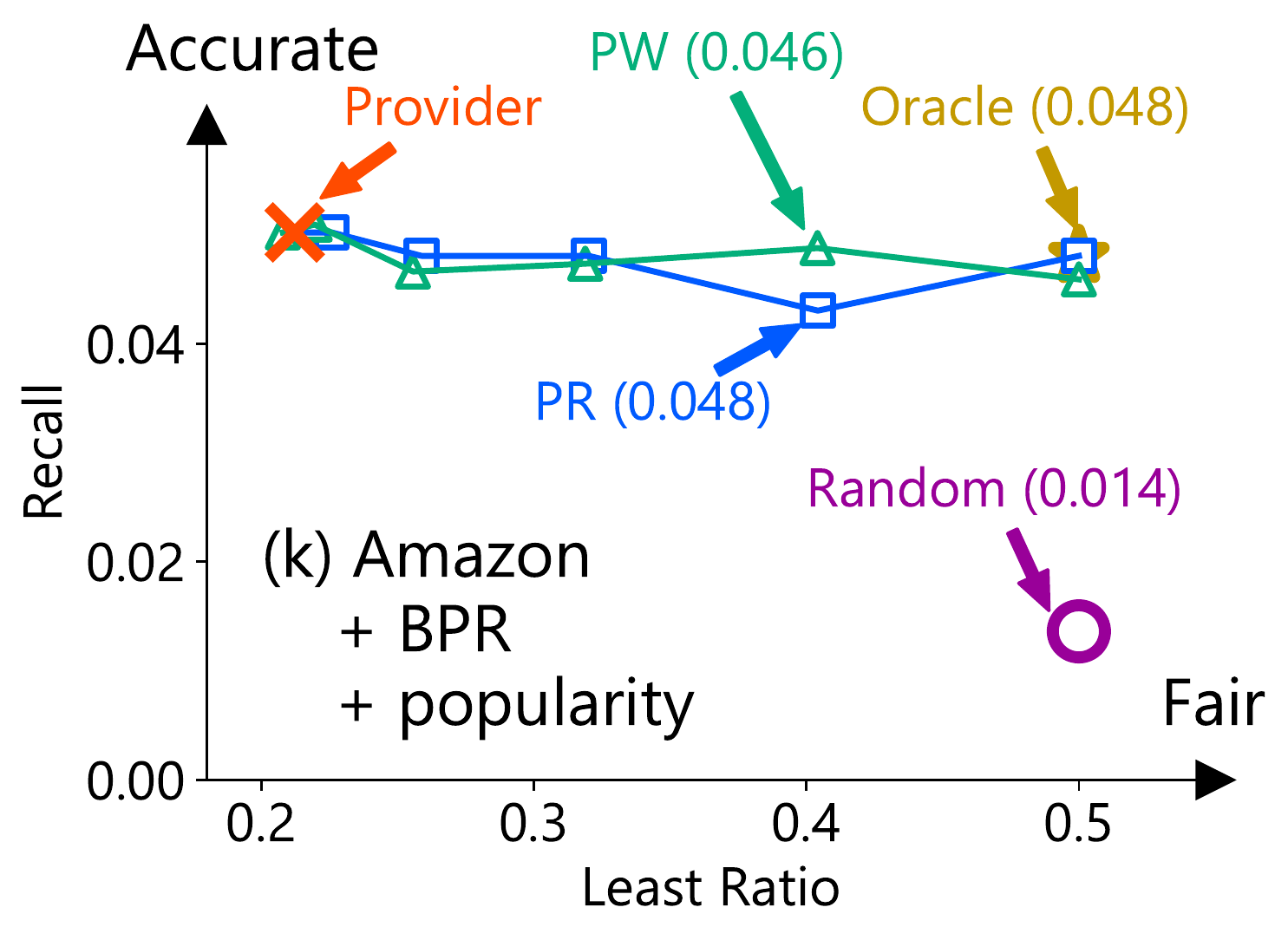}
\end{minipage}
\quad
\begin{minipage}{0.20\hsize}
\includegraphics[width=\hsize]{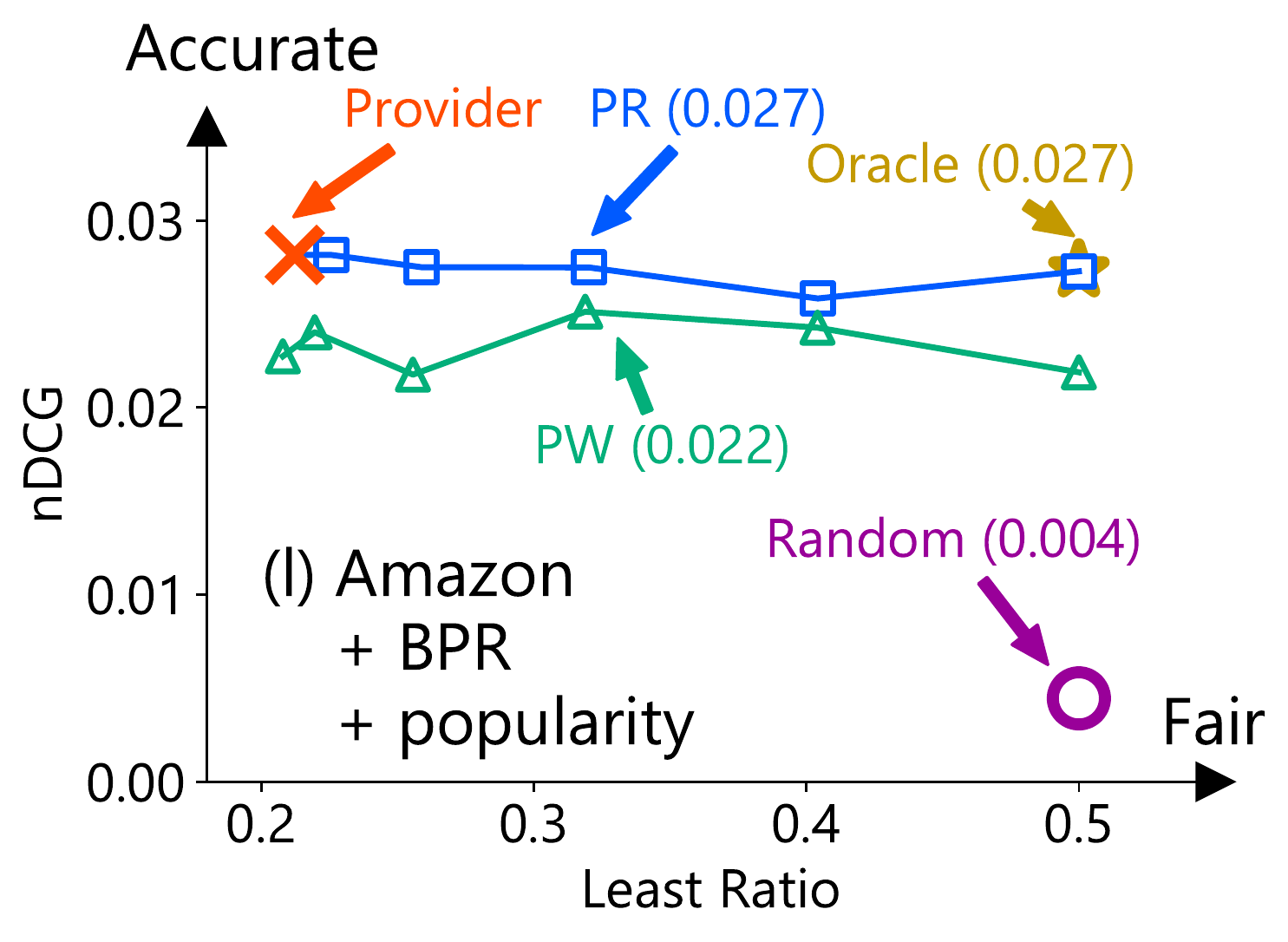}
\end{minipage}
\end{center}
\vspace{-0.1in}
\caption{Trade-off between fairness and performance: PR (blue curve) represents \textsc{PrivateRank}, and PW (green curve) represents \textsc{PrivateWalk}. The score reported in a parenthesis is the performance of each method \emph{when the recommendation is completely fair}. Even though \textsc{PrivateRank} does not use log data, its performance is comparable with the oracle method, which adopts prohibitive information. Although \textsc{PrivateWalk} performs worse than \textsc{PrivateRank} in exchange for fast evaluation, it remains significantly better than random guesses. }
\label{fig: result}
\vspace{-0.1in}
\end{figure*}

\subsubsection{Datasets}
We use four datasets, Adult \footnote{\url{https://archive.ics.uci.edu/ml/datasets/adult}}, MovieLens100k \cite{harper2016movielens}, LastFM \footnote{\url{https://grouplens.org/datasets/hetrec-2011/}}, and Amazon Home and Kitchen \cite{he2016ups, mcauley2015image}. In Adult, we recommend a set of people on each person's page, such that those recommended have the same label as the source person. A recommendation list is considered to be fair if it contains men and women in equal proportion. We use this dataset with talent searching in mind. In MovieLens, we consider two protected groups (i) movies released before 1990 and (ii) movies that received less than $50$ interactions. In LastFM and Amazon, items that received less than $50$ interactions are the protected groups. We adopt the implicit setting for MovieLens, LastFM, and Amazon datasets. We set the $ij$-th element of the interaction matrix to one if user $i$ has interacted with item $j$ and zero otherwise. More details about the datasets are available in Section \ref{sec: datasets} in the appendices.

\subsubsection{Provider Recommender System}
To construct private recommender systems, we first define the provider recommender system.

\noindent \textbf{Adult dataset.} We use a $K$-nearest neighbor recommendation for this dataset. It first standardizes features and recommends $K$-nearest people with respect to the Euclidean distance for each person.

\noindent \textbf{MovieLens, LastFM, and Amazon datasets.} We use nearest neighbor recommendation using rating vectors and Bayesian personalized ranking (BPR) \cite{rendle2009bpr}. In BPR, the similarity of items $i$ and $j$ is defined as the inner product of the latent vectors of items $i$ and $j$. The top-$K$ similar items with respect to these similarity measures are recommended in both methods. We use the Implicit package\footnote{\url{https://github.com/benfred/implicit}} to implement both methods.

\subsubsection{Evaluation Protocol} \hfill

\noindent \textbf{Adult dataset.} 
We evaluate methods using precision, i.e., the ratio of recommended items with the same class label as the source item.

\noindent \textbf{MovieLens and Amazon datasets.} We adopt leave-one-out evaluation following previous works \cite{he2017neural, rendle2009bpr}. We do not adopt negative sampling but full evaluation to avoid biased evaluation \cite{krichene2020sampled}. We evaluate methods using recall@$K$ and nDCG@$K$. These metrics are computed for the recommendation list for the second latest item that user $u$ interacted with. The positive sample is the latest interacted item, which is left out in the dataset.

\noindent \textbf{LastFM dataset.} This dataset does not contain timestamps. We randomly arrange the interactions and adopt leave-one-out evaluation as in MovieLens and Amazon datasets.

\subsubsection{Hyperparameters} Throughout the experiments, we set the length $K$ of recommendation lists to be $10$ for both provider and private recommender systems.
We set $c = 0.01$, $L = 10$ for \textsc{PrivateRank} and $L_\text{max} = 100$ for \textsc{PrivateWalk}. We inspect the trade-off between performance and fairness by varying $\tau$.
We use the default hyperparameters in the Implicit package for the BPR. Namely, the number of dimensions is $100$, the learning rate is $0.01$, the regularization parameter is $0.01$, and the number of iterations is $100$.

\subsection{Fairness and Performance Trade-off (RQ1)} \label{sec: result}

\subsubsection{Baselines} First, \textbf{Provider} is the provider recommender system. It serves as an upper bound of private recommender systems in terms of accuracy; however, it is unfair. \textbf{Random} shuffles items in random order and recommends items in a fair manner by the same post processing as \textsc{PrivateRank}. \textbf{Oracle} is a posthoc fair recommender system that adopts the same backbone algorithm as the provider recommender system and the same post processing as \textsc{PrivateRank}. This algorithm corresponds to the posthoc method in \cite{geyik2019fairness}. Note that this method is unrealistic and cannot be used in our setting because it does have access to log data. We use \textbf{Oracle} to investigate the performance deviations from idealistic settings. Note that existing methods for C-fairness, such as FATR \cite{zhu2018fairness} and Beyond Parity \cite{yao2017beyond}, cannot be used as baselines in this setting because we adopt fairness for \emph{items} in item-to-item recommendations in this work.

\subsubsection{Results}
Figure \ref{fig: result} shows the trade-off between the fairness measures and the performance of the recommender systems. The score reported in parenthesis is the performance of each method \emph{when the recommendation is completely fair} (i.e., least ratio is $0.5$ and entropy is $1.0$, not the score at the pointed position).

\begin{table*}[tb]
\vspace{-0.1in}
\small
    \caption{Case studies on IMDb and Twitter. \textsc{PrivateWalk} can retrieve relevant items in a fair manner, although it does not know the detail of the provider recommender system or have access to log data. Recall that we do not take orders into consideration in the fairness criterion. Once both groups are contained in a recommendation list, it is easy to adjust the order by any posthoc processing. (e.g., show the USA and non-USA movies alternatively in the first example if you want.)} 
    \vspace{-0.1in}
    \centering
    \scalebox{0.65}{
    \begin{tabular}{lllllll} \toprule
    & Provider (IMDb) & \textsc{PrivateWalk} & Provider (Twitter) & \textsc{PrivateWalk} & Provider (Twitter) & \textsc{PrivateWalk} \\ \midrule
    Source & Toy Story & Toy Story & Tom Hanks & Tom Hanks & Tim Berners-Lee & Tim Berners-Lee \\ \midrule
    1st & Toy Story 3 ({\color[HTML]{FF4B00} USA})    & Toy Story 3 ({\color[HTML]{FF4B00} USA})              & Jim Carrey ({\color[HTML]{FF4B00} man})        & Jim Carrey ({\color[HTML]{FF4B00} man})        & Jimmy Wales ({\color[HTML]{FF4B00} man})      & Vinton Gray Cerf ({\color[HTML]{FF4B00} man}) \\
    2nd & Toy Story 2 ({\color[HTML]{FF4B00} USA})    & Coco ({\color[HTML]{FF4B00} USA})                     & Hugh Jackman ({\color[HTML]{FF4B00} man})      & Sarah Silverman ({\color[HTML]{005AFF} woman}) & Vinton Gray Cerf ({\color[HTML]{FF4B00} man}) & Lawrence Lessig ({\color[HTML]{FF4B00} man}) \\
    3rd & Finding Nemo ({\color[HTML]{FF4B00} USA})   & The Incredibles ({\color[HTML]{FF4B00} USA})          & Samuel L. Jackson ({\color[HTML]{FF4B00} man}) & Hugh Jackman ({\color[HTML]{FF4B00} man})      & Nigel Shadbolt ({\color[HTML]{FF4B00} man})   & Nigel Shadbolt ({\color[HTML]{FF4B00} man}) \\
    4th & Monsters, Inc. ({\color[HTML]{FF4B00} USA}) & Spirited Away ({\color[HTML]{005AFF} non USA})        & Dwayne Johnson ({\color[HTML]{FF4B00} man})    & Samuel L. Jackson ({\color[HTML]{FF4B00} man}) & Anil Dash ({\color[HTML]{FF4B00} man})        & Kara Swisher ({\color[HTML]{005AFF} woman}) \\
    5th & Up ({\color[HTML]{FF4B00} USA})             & Castle in the Sky ({\color[HTML]{005AFF} non USA})    & Seth MacFarlane ({\color[HTML]{FF4B00} man})   & Emma Watson ({\color[HTML]{005AFF} woman})     & Lawrence Lessig ({\color[HTML]{FF4B00} man})  & danah boyd ({\color[HTML]{005AFF} woman}) \\
    6th & WALL·E ({\color[HTML]{FF4B00} USA})         & Howl's Moving Castle ({\color[HTML]{005AFF} non USA}) & Sarah Silverman ({\color[HTML]{005AFF} woman}) & Alyssa Milano ({\color[HTML]{005AFF} woman})   & Brendan Eich ({\color[HTML]{FF4B00} man})     & Adrienne Porter Felt ({\color[HTML]{005AFF} woman}) \\ \bottomrule
    \end{tabular}
    }
    \label{table: qualitative}
    \vspace{-0.15in}
\end{table*}

\noindent \textbf{Adult} (Figures \ref{fig: result} (a) and (b)): The least ratio of the provider recommendations is $0.152$, which indicates that the provider recommendations are not fair with respect to sex. In contrast, our proposed methods can increase fairness by increasing the threshold. In particular, \textsc{PrivateRank} strikes an excellent trade-off between fairness measures and precision. It achieves perfect fairness (i.e., least ratio is $0.5$ and entropy is $1.0$) while it drops precision by only one percent. \textsc{PrivateWalk} performs slightly worse than \textsc{PrivateRank} but much better than random access. Because the least ratio and entropy have a one-to-one correspondence, we will report only the least ratios in the following because of space limitation.

\noindent \textbf{MovieLens} (Figures \ref{fig: result} (c) to (h)): Cosine and BPR in the figure represent the type of the provider recommender system, and popularity and period represent the protected attributes. \textsc{PrivateRank} strikes an excellent trade-off between accuracy and fairness in all settings. In particular, \textsc{PrivateRank} is comparable with the oracle method, which has access to the unavailable information in our setting. It can also be observed that BPR increases the performance of the provider recommender system compared to the cosine similarity, and it also increases the performance of private recommender systems accordingly. This indicates that effective provider recommender systems induce effective private recommender systems. The overall tendencies are common in all settings. These results indicate that our proposed methods can generate effective recommendations in a fair manner regardless of the algorithms used in the provider recommender system and the protected attributes. Because we observe similar tendencies in Cosine and BPR for other datasets, we report only BPR in the following due to space limitation.

\noindent \textbf{LastFM and Amazon} (Figures \ref{fig: result} (i) to (l)): Our proposed methods exhibits good trade-off between accuracy and fairness here as well. These results indicate that our proposed methods are effective regardless of the domain of recommendations.

In summary, our proposed methods can achieve perfect fairness, even where the provider recommender systems are not fair, and perform well in various settings.

\subsection{Sensitivity of Hyperparameters (RQ2)} \label{sec: sensitivity}

\begin{figure}[tb]
\begin{minipage}{0.32\hsize}
\centering
\includegraphics[width=\hsize]{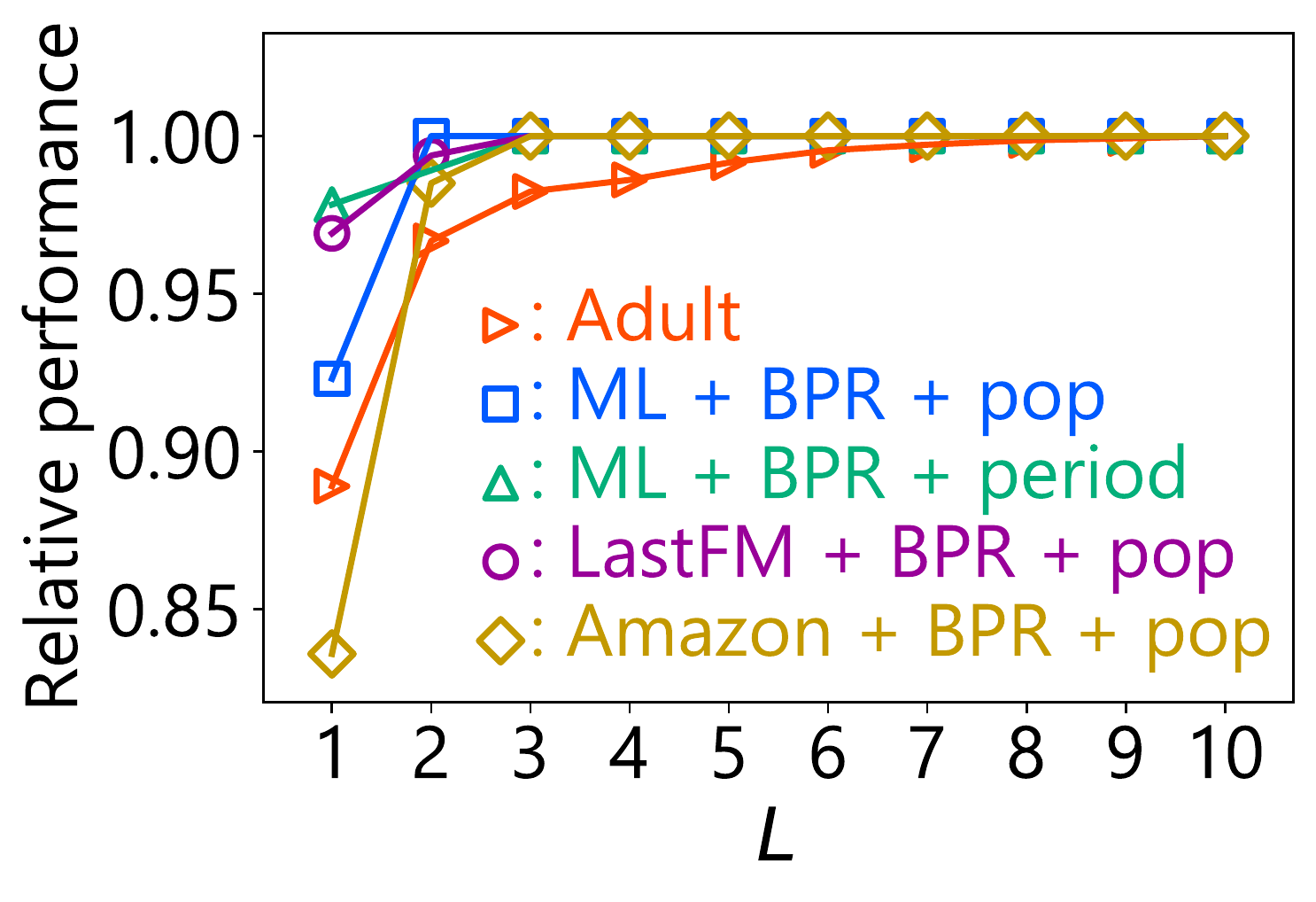}
\end{minipage}
\begin{minipage}{0.32\hsize}
\centering
\includegraphics[width=\hsize]{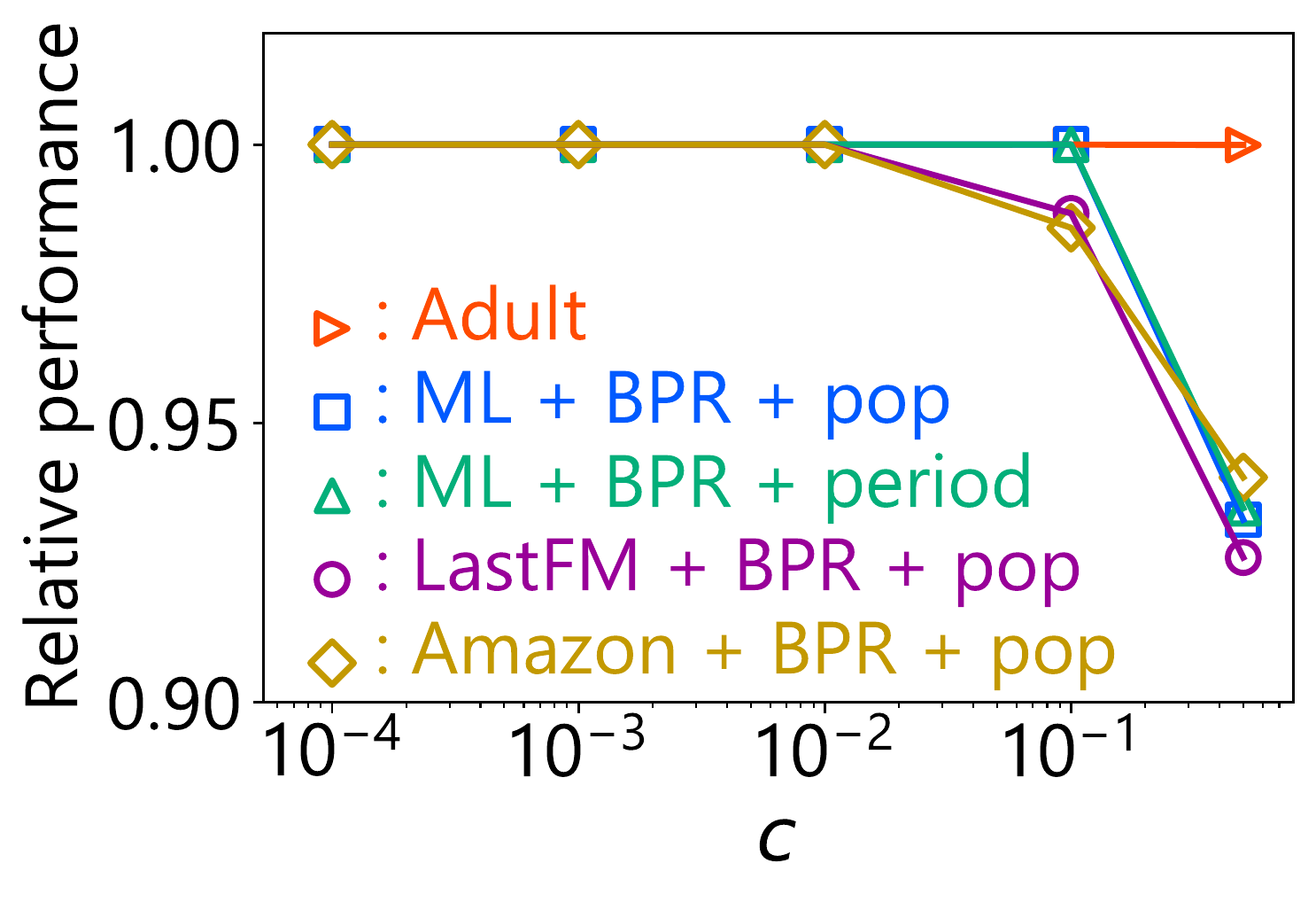}
\end{minipage}
\begin{minipage}{0.32\hsize}
\centering
\includegraphics[width=\hsize]{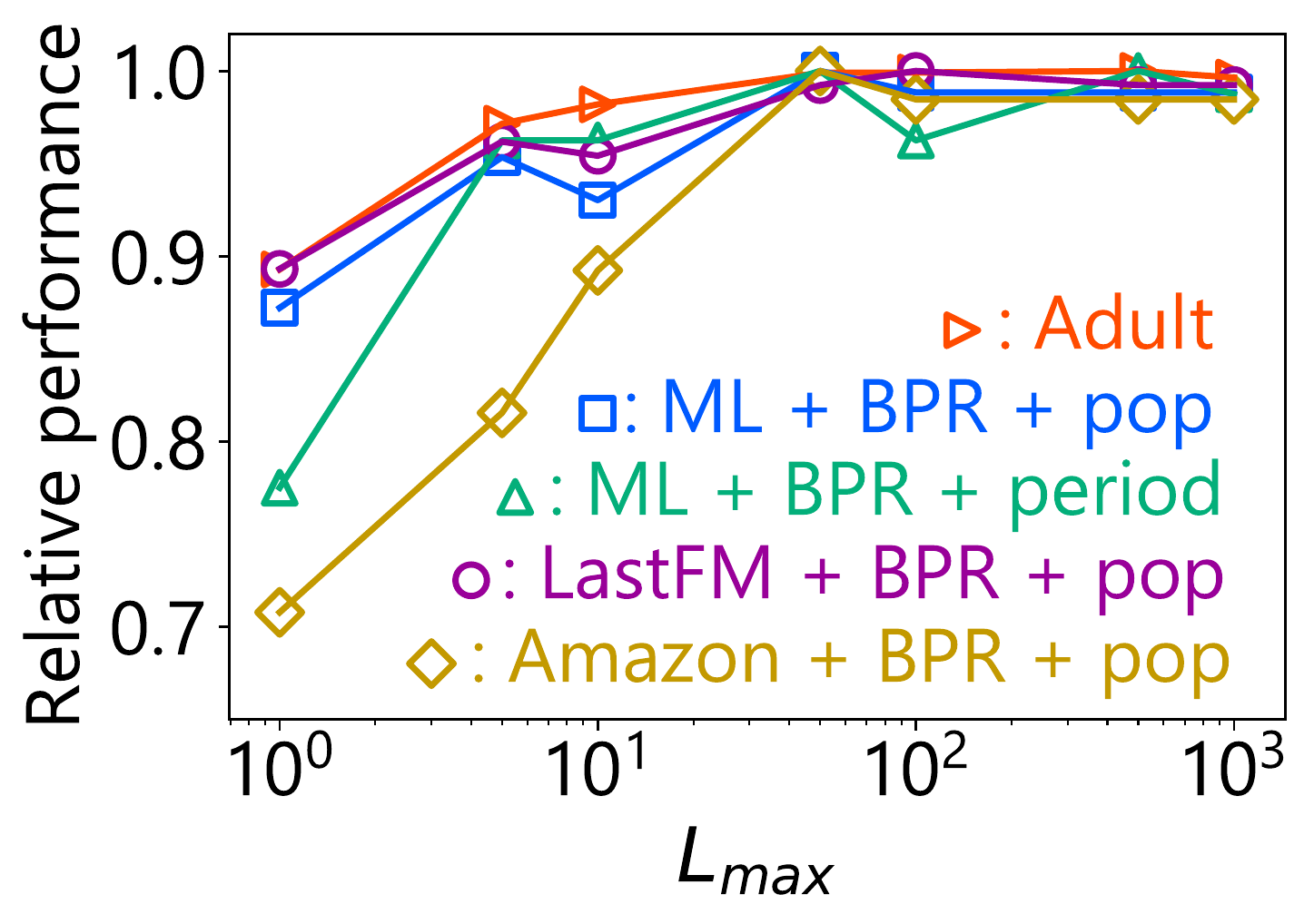}
\end{minipage}
\vspace{-0.1in}
\caption{Sensitivity of hyperparameters: Our methods are insensitive to hyperparameters.}
\label{fig: sensitivity}
\vspace{-0.2in}
\end{figure}

We investigate the sensitivity of hyperparameters of \textsc{PrivateRank} and \textsc{PrivateWalk}. We fix the minimum requirement $\tau$ to be $5$ (i.e., perfect fairness) and evaluate performance with various hyperparameters. We evaluate performance by precision for the Adult dataset and by the recall for the other datasets. We normalize these performance measures such that the maximum value is one to illustrate relative drops of performance. Figure \ref{fig: sensitivity} (Left) reports the sensitivity of the number $L$ of iterations of the cumulative power iteration, and Figure \ref{fig: sensitivity} (Center) reports the sensitivity of the damping factor $c$. It can be observed that $L$ is not sensitive as long as $L \ge 6$, and $c$ is neither sensitive as long as $c \le 0.01$ in all settings. Figure \ref{fig: sensitivity} (Right) reports the sensitivity of the maximum length $L_\text{max}$ of random walks. It also shows that this hyperparameter is not sensitive as long as $L_\text{max} \ge 100$. 

\subsection{Case Studies in the Wild (RQ3)} \label{sec: wild}

We run \textsc{PrivateWalk} with real recommender systems in operation in IMDb and Twitter for qualitative case studies. The main goal of this section is to show that our method is feasible even in real-world environments. To the best of our knowledge, there are no existing methods that enable Twitter users to use fair recommender systems. The experiments in this section provide a proof-of-concept that such a challenging task is feasible.

\noindent \textbf{IMDb:} We found that IMDb recommended only American movies in the Toy Story page\footnote{\url{https://www.imdb.com/title/tt0114709/}}. Although these recommendations are consistent, they are not informative to cinema fans. We consider whether the movie is from the USA as the protected attribute and run \textsc{PrivateWalk} on IMDb. Table \ref{table: qualitative} (Left) reports that \textsc{PrivateWalk} also recommends Spirited Away, Castle in the Sky, and Howl's Moving Castle, which are Japanese animation movies. This result indicates that \textsc{PrivateWalk} recommends in a fair manner with respect to the country attribute while it recommends relevant items.

\noindent \textbf{Twitter:} As we mentioned in Section \ref{sec: setting}, we found that Twitter's user recommendations were not fair with respect to gender, which is in contrast to LinkedIn. We set the protected attribute to be gender and run \textsc{PrivateWalk} on Twitter. We annotate labels and remove non-person accounts manually\footnote{As we mentioned in Section \ref{sec: setting}, this annotation process can be automated if there is a classifier that takes an account as input and estimates the label.}. Table \ref{table: qualitative} (Middle, Right) reports that \textsc{PrivateWalk} recommends three men users and three woman users in both examples, and all recommended users are celebrity actors for Tom Hanks, and tech-related people for Tim Berners-Lee.

We emphasize that we do not know the recommendation algorithms used in IMDb and Twitter nor have access to log data. Nonetheless, \textsc{PrivateWalk} generates fair recommendations by utilizing blackbox recommendation results. We also point out that \textsc{PrivateRank} is infeasible in this setting because there are too many items in IMDb and Twitter to hit the API limit. In contrast, \textsc{PrivateWalk} runs with few evaluations of the provider recommender systems in an on-demand manner. \textsc{PrivateRank} is beneficial when the number of items is small (e.g., when we retrieve items only from a specific category) or when we can crawl the website beforehand because it performs better when allowed.

\section{Discussion}
\noindent \uline{\textbf{Limitation.}} The main limitation of private recommender systems is that it is expensive for each user to develop a system. Although it may still be worthwhile for companies to build their own private recommender systems for a fair recruiting process, most individual users cannot afford to develop their own browser add-ons. We consider two specific scenarios in which individual users benefit from private recommender systems. First, some enthusiastic users of the service can develop browser add-ons and distribute them. For example, a cinema fan may develop a fair private recommender system for IMDb with respect to popularity as a hobby project, such that other IMDb users can enjoy it. Second, general purpose software may help build private recommender systems. In particular, the program we developed in Section \ref{sec: wild} computes random walks and rankings in common procedures, and we only have to specify in which elements of a web page recommendation lists and sensitive attributes are described. Although users have to write additional scripts for each service in the current version, more elaborated software may further reduce the burden of developing private recommender systems.

We have focused on item-to-item collaborative filtering, which is one of the popular choices in steerable recommender systems \cite{green2009generating, lamprecht2015improving}. Extending the concept of private recommender systems to other settings, including item-to-user setting, would be fruitful future direction.

\vspace{0.05in}
\noindent \uline{\textbf{Extending recommendation lists.}} So far, we have assumed that the length of the recommendation lists of the private recommender system is the same as that of the provider recommender system. However, some users may want to know more items than the service provider offers. Our proposed methods can drop this assumption and provide longer lists than the service provider.

\section{Related Work}

Burke \cite{bruke2017mltisided} classified fairness of recommender systems into three categories. C-fairness is the fairness of consumers or users of a service, P-fairness is the fairness of producers or items, and CP-fairness considers both sides. In this study, we focused on P-fairness. Note that existing methods for C-fairness such as FATR \cite{zhu2018fairness} cannot be applied in this setting. Examples of previous studies on P-fairness are as follows. Geyik et al. \cite{geyik2019fairness} proposed a fair ranking of job-seekers in LinkedIn Talent Search. Ekstrand et al. \cite{ekstrand2018exploring} studied book recommendations and found that some recommendation algorithms are biased toward books written by men authors. Beutel et al. \cite{beutel2019fairness} were concerned that unfair recommendations in social network services might under-rank posts by some demographic groups, which limits the groups' visibility. Mehrotra et al. \cite{mehrotra2018towards} and Patro et al. \cite{patro2020fairrec} proposed methods to equalize the visibility of items and realize a fair marketplace. The crucial difference between these studies and our research is that all of these previous methods are for service providers and require log data. In contrast, our methods can be used even without log data and even if a service provider does not offer a fair recommender system. To the best of our knowledge, this is the first work to address this challenging problem.

\section{Conclusion}

In this paper, we investigated a situation where a service provider does not offer a fair recommender system for the first time. We proposed a novel framework, private recommender systems, where each user builds their own fair recommender system. We proposed \textsc{PrivateRank} and \textsc{PrivateWalk} to build a private recommender system without requiring access to log data. \textsc{PrivateRank} is effective and exhibits an excellent trade-off between performance and fairness; however, it requires many evaluations of the provider recommender system. Although \textsc{PrivateWalk} is less effective, it makes recommendations in an on-demand manner and requires few evaluations of the provider recommender system. The two proposed methods complement each other’s weaknesses. We empirically validated the effectiveness of the proposed methods via offline quantitative evaluations and qualitative experiments in the wild. Our approach realizes fair recommendations in many cases where conventional fair algorithms cannot be deployed, and it promotes the spread of fair systems. 

\bibliographystyle{siam}
\bibliography{sample-base}

\begin{thebibliography}{10}

\bibitem{beutel2019fairness}
{\sc A.~Beutel, J.~Chen, T.~Doshi, H.~Qian, L.~Wei, Y.~Wu, L.~Heldt, Z.~Zhao,
  L.~Hong, E.~H. Chi, and C.~Goodrow}, {\em Fairness in recommendation ranking
  through pairwise comparisons}, in {KDD}, 2019, pp.~2212--2220.

\bibitem{bruke2017mltisided}
{\sc R.~Burke}, {\em Multisided fairness for recommendation}, in {FAT/ML},
  2017.

\bibitem{cano2006topology}
{\sc P.~Cano, O.~Celma, M.~Koppenberger, and J.~M. Buldu}, {\em Topology of
  music recommendation networks}, Chaos: An interdisciplinary journal of
  nonlinear science, 16 (2006), p.~013107.

\bibitem{celma2008new}
{\sc {\`{O}}.~Celma and P.~Herrera}, {\em A new approach to evaluating novel
  recommendations}, in RecSys, 2008, pp.~179--186.

\bibitem{corbett2017algorithmic}
{\sc S.~Corbett{-}Davies, E.~Pierson, A.~Feller, S.~Goel, and A.~Huq}, {\em
  Algorithmic decision making and the cost of fairness}, in {SIGKDD}, 2017,
  pp.~797--806.

\bibitem{dua2017uci}
{\sc D.~Dua and C.~Graff}, {\em {UCI} machine learning repository}, 2017.

\bibitem{ekstrand2018exploring}
{\sc M.~D. Ekstrand, M.~Tian, M.~R.~I. Kazi, H.~Mehrpouyan, and D.~Kluver},
  {\em Exploring author gender in book rating and recommendation}, in RecSys,
  2018, pp.~242--250.

\bibitem{feldman2015certifying}
{\sc M.~Feldman, S.~A. Friedler, J.~Moeller, C.~Scheidegger, and
  S.~Venkatasubramanian}, {\em Certifying and removing disparate impact}, in
  {KDD}, 2015, pp.~259--268.

\bibitem{geyik2019fairness}
{\sc S.~C. Geyik, S.~Ambler, and K.~Kenthapadi}, {\em Fairness-aware ranking in
  search {\&} recommendation systems with application to linkedin talent
  search}, in {KDD}, 2019, pp.~2221--2231.

\bibitem{geyik2018building}
{\sc S.~C. Geyik and K.~Kenthapadi}, {\em Building representative talent search
  at {LinkedIn}}, 2018.

\bibitem{green2009generating}
{\sc S.~J. Green, P.~Lamere, J.~Alexander, F.~Maillet, S.~Kirk, J.~Holt,
  J.~Bourque, and X.~Mak}, {\em Generating transparent, steerable
  recommendations from textual descriptions of items}, in Proceedings of the
  2009 {ACM} Conference on Recommender Systems, RecSys, {ACM}, 2009,
  pp.~281--284.

\bibitem{harper2016movielens}
{\sc F.~M. Harper and J.~A. Konstan}, {\em The movielens datasets: History and
  context}, {ACM} Trans. Interact. Intell. Syst., 5 (2016), pp.~19:1--19:19.

\bibitem{he2016ups}
{\sc R.~He and J.~J. McAuley}, {\em Ups and downs: Modeling the visual
  evolution of fashion trends with one-class collaborative filtering}, in
  {WWW}, 2016, pp.~507--517.

\bibitem{he2017neural}
{\sc X.~He, L.~Liao, H.~Zhang, L.~Nie, X.~Hu, and T.~Chua}, {\em Neural
  collaborative filtering}, in {WWW}, 2017, pp.~173--182.

\bibitem{jeh2003scaling}
{\sc G.~Jeh and J.~Widom}, {\em Scaling personalized web search}, in {WWW},
  2003, pp.~271--279.

\bibitem{kamishima2012fairness}
{\sc T.~Kamishima, S.~Akaho, H.~Asoh, and J.~Sakuma}, {\em Fairness-aware
  classifier with prejudice remover regularizer}, in {ECML} {PKDD}, vol.~7524,
  2012, pp.~35--50.

\bibitem{koren2009matrix}
{\sc Y.~Koren, R.~M. Bell, and C.~Volinsky}, {\em Matrix factorization
  techniques for recommender systems}, Computer, 42 (2009), pp.~30--37.

\bibitem{krichene2020sampled}
{\sc W.~Krichene and S.~Rendle}, {\em On sampled metrics for item
  recommendation}, in {KDD}, 2020, pp.~1748--1757.

\bibitem{lamprecht2015improving}
{\sc D.~Lamprecht, F.~Geigl, T.~Karas, S.~Walk, D.~Helic, and M.~Strohmaier},
  {\em Improving recommender system navigability through diversification: a
  case study of imdb}, in Proceedings of the 15th International Conference on
  Knowledge Technologies and Data-driven Business, {I-KNOW}, {ACM}, 2015,
  pp.~21:1--21:8.

\bibitem{liu2019personalized}
{\sc W.~Liu, J.~Guo, N.~Sonboli, R.~Burke, and S.~Zhang}, {\em Personalized
  fairness-aware re-ranking for microlending}, in RecSys, {ACM}, 2019,
  pp.~467--471.

\bibitem{mcauley2015image}
{\sc J.~J. McAuley, C.~Targett, Q.~Shi, and A.~van~den Hengel}, {\em
  Image-based recommendations on styles and substitutes}, in {SIGIR}, 2015,
  pp.~43--52.

\bibitem{mehrotra2018towards}
{\sc R.~Mehrotra, J.~McInerney, H.~Bouchard, M.~Lalmas, and F.~Diaz}, {\em
  Towards a fair marketplace: Counterfactual evaluation of the trade-off
  between relevance, fairness {\&} satisfaction in recommendation systems}, in
  {CIKM}, 2018, pp.~2243--2251.

\bibitem{milano2020recommender}
{\sc S.~Milano, M.~Taddeo, and L.~Floridi}, {\em Recommender systems and their
  ethical challenges}, {AI} Soc., 35 (2020), pp.~957--967.

\bibitem{page1999pagerank}
{\sc L.~Page, S.~Brin, R.~Motwani, and T.~Winograd}, {\em The pagerank citation
  ranking: Bringing order to the web.}, tech. rep., Stanford InfoLab, 1999.

\bibitem{pariser2011filter}
{\sc E.~Pariser}, {\em The filter bubble: What the Internet is hiding from
  you}, Penguin UK, 2011.

\bibitem{patro2020fairrec}
{\sc G.~K. Patro, A.~Biswas, N.~Ganguly, K.~P. Gummadi, and A.~Chakraborty},
  {\em Fairrec: Two-sided fairness for personalized recommendations in
  two-sided platforms}, in {WWW}, 2020, pp.~1194--1204.

\bibitem{rendle2009bpr}
{\sc S.~Rendle, C.~Freudenthaler, Z.~Gantner, and L.~Schmidt{-}Thieme}, {\em
  {BPR:} bayesian personalized ranking from implicit feedback}, in {UAI}, 2009,
  pp.~452--461.

\bibitem{seyerlehner2009limitation}
{\sc K.~Seyerlehner, A.~Flexer, and G.~Widmer}, {\em On the limitations of
  browsing top-n recommender systems}, in RecSys, 2009, pp.~321--324.

\bibitem{seyerlehner2009browsing}
{\sc K.~Seyerlehner, P.~Knees, D.~Schnitzer, and G.~Widmer}, {\em Browsing
  music recommendation networks}, in Proceedings of the 10th International
  Society for Music Information Retrieval Conference, {ISMIR}, International
  Society for Music Information Retrieval, 2009, pp.~129--134.

\bibitem{singh2018fairness}
{\sc A.~Singh and T.~Joachims}, {\em Fairness of exposure in rankings}, in
  {KDD}, 2018, pp.~2219--2228.

\bibitem{xiao2019beyond}
{\sc W.~Xiao, H.~Zhao, H.~Pan, Y.~Song, V.~W. Zheng, and Q.~Yang}, {\em Beyond
  personalization: Social content recommendation for creator equality and
  consumer satisfaction}, in {KDD}, 2019, pp.~235--245.

\bibitem{xu2020algorithmic}
{\sc R.~Xu, P.~Cui, K.~Kuang, B.~Li, L.~Zhou, Z.~Shen, and W.~Cui}, {\em
  Algorithmic decision making with conditional fairness}, in {KDD}, 2020,
  pp.~2125--2135.

\bibitem{yao2017beyond}
{\sc S.~Yao and B.~Huang}, {\em Beyond parity: Fairness objectives for
  collaborative filtering}, in {NeurIPS}, 2017, pp.~2921--2930.

\bibitem{yoon2018tpa}
{\sc M.~Yoon, J.~Jung, and U.~Kang}, {\em {TPA:} fast, scalable, and accurate
  method for approximate random walk with restart on billion scale graphs}, in
  {ICDE}, 2018, pp.~1132--1143.

\bibitem{zehlike2017fair}
{\sc M.~Zehlike, F.~Bonchi, C.~Castillo, S.~Hajian, M.~Megahed, and
  R.~Baeza{-}Yates}, {\em {FA*IR:} {A} fair top-k ranking algorithm}, in
  {CIKM}, 2017, pp.~1569--1578.

\bibitem{zhu2018fairness}
{\sc Z.~Zhu, X.~Hu, and J.~Caverlee}, {\em Fairness-aware tensor-based
  recommendation}, in Proceedings of the 27th {ACM} International Conference on
  Information and Knowledge Management, {CIKM}, 2018, pp.~1153--1162.

\bibitem{ziegler2005improving}
{\sc C.~Ziegler, S.~M. McNee, J.~A. Konstan, and G.~Lausen}, {\em Improving
  recommendation lists through topic diversification}, in {WWW}, 2005,
  pp.~22--32.

\end{thebibliography}

\newpage

\appendix

\section{Notations}

Notations are summarized in Table \ref{tab: notations}.

\begin{table}[tb]
    \centering
    \caption{Notations.}
    \begin{tabular}{ll} \toprule
        Notations & Descriptions \\ \midrule
        $[n]$ & The set $\{ 1, 2, \dots, n \}$. \\
        $a, \bolda, \boldA$ & A scalar, vector, and matrix. \\
        $\boldA^\top$ & The transpose of $\boldA$. \\
        $K$ & The length of a recommendation list. \\
        $\mathcal{U} = [m]$ & The set of users. \\
        $\mathcal{I} = [n]$ & The set of items. \\
        $\mathcal{A}$ & The set of protected groups. \\
        $a_i \in \mathcal{A}$ & The protected attribute of item $i \in \mathcal{I}$. \\
        $\mathcal{P}_{\text{prov}}$ & A provider recommender system. \\
        $\mathcal{Q}$ & A private recommender system. \\
        \bottomrule
    \end{tabular}
    \label{tab: notations}
\end{table}

\section{Pseudo Code of PrivateRank} \label{sec: PrivateRankcode}

\setlength{\textfloatsep}{5pt}
\begin{algorithm2e}[t]
\caption{CanAdd$(\mathcal{R}, i)$}
\label{algo: can}
\DontPrintSemicolon 
\nl\KwData{List $\mathcal{R}$ of items, New item $i$.}
\nl\KwResult{Whether we can add item $i$.}
    \nl Initialize $\boldc_a \leftarrow 0 ~(\forall a \in \mathcal{A}$) \;
    \nl \For{$j \textup{ \textbf{in} } \mathcal{R} \cup \{i\}$}{
    \nl     $\boldc_{a_j} \leftarrow \boldc_{a_j} + 1$ \tcp*{Count attributes}
        }
    \nl \textbf{return} $\sum_{a \in \mathcal{A}} \max(0, \tau - c_a) \le K - \text{len}(\mathcal{R})$ \;
\end{algorithm2e}

Algorithm \ref{algo: PrivateRank} describes the pseudo-code of \textsc{PrivateRank}. We construct the recommendation network in Line 4 and compute the personalized PageRank in Line 5. In Lines 6--7, we iterate items in the descending order of the personalized PageRank. In Lines 8--9, we add item $i$ to the recommendation list if the user has not interacted with item $i$, and we can preserve the constraint when we insert item $i$.

\section{Proof of Theorem 4.1} \label{sec: thm1}

\begin{proof}
Let $o_i \in [n]$ be the rank of item $i$ in $\hat{\boldS}_s$. Item $i$ appears in the $o_i$-th iteration, and item ord$_i$ appears in the $i$-th iteration of the loop in Lines 7--11 in Algorithm \ref{algo: PrivateRank}. We prove that for each iteration $j \in [n]$ and each sensitive attribute $a \in \mathcal{A}$, $|\{i \in \mathcal{R} \mid a_i = a\} \cup \{i \in \mathcal{I} \mid a_i = a \land o_i \ge j\}| \ge \tau$ holds at the start of the $j$-th iteration. We prove this assertion by mathematical induction. The $j = 1$ case holds because there are at least $\tau$ items of each attribute. Suppose the proposition holds for $j$. Let $\hat{a} = a_{\text{ord}_j}$. The proposition holds for sensitive attribute $a \neq \hat{a}$ because neither $\{i \in \mathcal{R} \mid a_i = a\}$ nor $\{i \in \mathcal{I} \mid a_i = a \land o_i \ge j\}$ changes in the $j$-th iteration. If $|\{i \in \mathcal{R} \mid a_i = \hat{a}\}| \ge \tau$ holds in the $j$-th iteration, the proposition holds for $\hat{a}$. Otherwise, item $\text{ord}_j$ is adopted in the $j$-th iteration because $\sum_{a \in \mathcal{A}} \max(0, \tau - c_a)$ decreases by one, and CanAdd returns \textbf{True}. Therefore, $|\{i \in \mathcal{R} \mid a_i = a\}|$ increases by one, and $|\{i \in \mathcal{I} \mid a_i = a \land o_i \ge j\}|$ decreases by one at the $j$-th iteration. From the inductive hypothesis, the proposition holds for $j + 1$. Therefore, $|\{i \in \mathcal{R} \mid a_i = a\}| \ge \tau$ holds true at the end of the procedure for each attribute $a \in \mathcal{A}$. 
\end{proof}

\section{Proof of Theorem 4.2} \label{sec: thm2}

\begin{proof}
Let $D = \sum_{i=1}^K \frac{1}{\log (i + 1)}$. From the definition,
\begin{align*}
    \hat{\boldS}_i = (1-c) \bolde^{(i)} + (1-c)c \left( \tilde{\boldA}^\top \bolde^{(i)} + c\tilde{\boldA}^\top \sum_{k = 0}^{L-2} (c \tilde{\boldA}^\top)^k \bolde^{(i)} \right).
\end{align*}
Therefore, for the $k$-th item $j$ of $\mathcal{P}_{\text{prov}, 1}(i)$, 
\[\hat{\boldS}_{ij} \ge (1-c)c (\tilde{\boldA}^\top \bolde^{(i)})_j = (1-c)c \frac{1}{D \log (k + 1)}.\]
Suppose $c < \frac{1}{(K + 1)^2 \log^2(K + 1)}$ holds, then, 
\begin{align*}
    & \|c \tilde{\boldA}^\top \sum_{k = 0}^{L - 2} (c \tilde{\boldA}^\top)^k \bolde^{(i)}\|_\infty \\
    &\le c \|\tilde{\boldA}^\top\|_1 \|\sum_{k = 0}^{\infty} (c \tilde{\boldA}^\top)^k\|_1 \|\bolde^{(i)}\|_1 \\
    &\le c \|\tilde{\boldA}^\top\|_1 \frac{1}{1 - \|c \tilde{\boldA}^\top\|_1} \|\bolde^{(i)}\|_1 \\
    &\le \frac{c}{1 - c} = \frac{1}{(K + 1)^2 \log^2(K + 1) - 1} \\
    &< \frac{1}{D(K+1)\log^2(K + 1)} < \frac{\left( \frac{1}{\log K} - \frac{1}{\log (K + 1)} \right)}{D}, \\
\end{align*}
where $\|\bolda\|_1$ is the induce norm. Therefore, for $l \not \in \mathcal{P}_{\text{prov}, 1}(i) \cup \{i\}$, 
\[ \hat{\boldS}_{il} = (1-c)c (c\tilde{\boldA}^\top \sum_{k = 0}^{L-2} (c \tilde{\boldA}^\top)^k \bolde^{(i)}) < (1-c)c \frac{1}{D\log (K + 1)},\] and the $k$-th item $j$ of $\mathcal{P}_{\text{prov}, 1}(i)$, 
\[\hat{\boldS}_{ij} < (1-c)c \frac{1}{D\log (k)}.\]
Therefore, the top-$K$ ranking does not change.

\end{proof}

\section{Pseudo Code of PrivateWalk} \label{sec: PrivateWalkcode}

\setlength{\textfloatsep}{5pt}
\begin{algorithm2e}[t]
\caption{\textsc{PrivateRank}}
\label{algo: PrivateRank}
\DontPrintSemicolon 
\nl\KwData{Oracle access to $\mathcal{P}_{\text{prov}, 1}$, Source item $i \in \mathcal{I}$, Protected attributes $a_i ~\forall i \in \mathcal{I}$, Damping factor $c$, Minimum requirement $\tau$, Set $\mathcal{H}$ of items that user $i$ has already interacted with.}
\nl\KwResult{Recommended items $\mathcal{R} = \{j_k\}_{1 \le k \le K}$.}
    \nl Initialize $\mathcal{R} \leftarrow []$ (empty) \;
    \nl $\boldA_{ij} = \begin{cases}
    \frac{1}{\log(k + 1)} & (\mathcal{P}_{\text{prov}, 1}(i)_k = j) \\
    0 & \text{(otherwise)}
    \end{cases}$\;
    \nl $\hat{\boldS}_{i} = (1 - c) \sum_{k = 0}^L (c \tilde{\boldA}^\top)^k \bolde^{(i)}$ \;
    \nl $l \leftarrow \text{argsort}(\hat{\boldS}_{i}, \text{descending})$ \;
    \nl \For{$i \textup{ \textbf{in} } l$}{
    \nl     \If{$i \textup{ \textbf{not in} } \mathcal{H} \textup{ \textbf{and} } \textup{CanAdd}(\mathcal{R}, i)$}{
    \nl         Push back $i$ to $\mathcal{R}$ \;
            }
        }
    \nl \textbf{return} $\mathcal{R}$ \;
\end{algorithm2e}

\setlength{\textfloatsep}{5pt}
\begin{algorithm2e}[t]
\caption{\textsc{PrivateWalk}}
\label{algo: PrivateWalk}
\DontPrintSemicolon 
\nl\KwData{Oracle access to $\mathcal{P}_{\text{prov}, 1}$, Source item $i \in \mathcal{I}$, Protected attributes $a_i ~\forall i \in \mathcal{I}$, Minimum requirement $\tau$, Set $\mathcal{H}$ of items that user $i$ has already interacted with, Maximum length $L_{\text{max}}$ of random walks.}
\nl\KwResult{Recommended items $\mathcal{R} = \{j_k\}_{1 \le k \le K}$.}
    \nl Initialize $\mathcal{R} \leftarrow []$ (empty) \;
    \nl \For{$k \gets 1$ \textbf{to} $K$}{
    \nl     \text{cur} $\leftarrow i$ \tcp*{start from the source node}
    \nl     \text{found} $\leftarrow$ \textbf{False} \;
    \nl     \For{\textup{iter} $\gets 1$ \textbf{to} $L_{\text{max}}$}{
    \nl         \text{next\_rank} $\sim$ $\text{Cat}(\text{normalize}([1 / \log (r + 1)]))$ \;
    \nl         \text{cur} $\leftarrow \mathcal{P}_{\text{prov}, 1}(\text{cur})_{\text{next\_rank}}$ \;
    \nl         \If{$\textup{cur} \textup{ \textbf{not} \textbf{in} } \mathcal{R} \cup \mathcal{H}$ \textup{\textbf{and}} $\textup{CanAdd}(\mathcal{R}, \textup{cur})$}{
    \nl             \tcc{The first encountered item that can be added. Avoid items in $\mathcal{R} \cup \mathcal{H}$.}
    \nl             Push back \text{cur} to $\mathcal{R}$. \;
    \nl             \text{found} $\leftarrow$ \textbf{True} \;
    \nl             \textbf{break} \;
                }
            }
    \nl     \While{\textbf{\textup{not}} \textup{found}}{
    \nl         $i \leftarrow$ \text{Uniform}($\mathcal{I}$) \tcp*{random item}
    \nl         \If{$i \textup{ \textbf{not} \textbf{in} } \mathcal{R} \cup \mathcal{H}$ \textup{\textbf{and}} $\textup{CanAdd}(\mathcal{R}, i)$}{
    \nl             Push back \text{cur} to $\mathcal{R}$. \;
    \nl             \textbf{break} \;
                }
            }
        }
    \nl \textbf{return} $\mathcal{R}$ \;
\end{algorithm2e}

Algorithm \ref{algo: PrivateWalk} presents the pseudo-code of \textsc{PrivateWalk}. $\text{Cat}(\theta)$ denotes a categorical distribution of parameter $\theta$, and $\text{normalize}(x)$ denotes the normalization of vector $x$, such that the sum is equal to one. We run a random walk in Lines 7--14. In Line 8, the next node is sampled with a probability proportional to edge weights. If we cannot find an appropriate item in $L_\text{max}$ steps, we add a random item in Lines 15--19. Note that we can determine an item in Lines 7--14 with high probability in practice. We add the fallback process in Lines 15--19 to make the algorithm well-defined.

\section{Datasets} \label{sec: datasets}

\begin{itemize}
    \item \textbf{Adult} \cite{dua2017uci}. In this dataset, each record represents a person and contains demographic data such as age, sex, race, educational background, and income. This dataset has been mainly adopted in supervised learning \cite{kamishima2012fairness, xu2020algorithmic}, where the covariates are demographic attributes, and the label represents whether income exceeds \$50,000 per year. We use this dataset for individual recommendations. We regard an individual record as a talent and recommend a set of people on each person's page, such that those recommended have the same label as the source person. A recommendation list is considered to be fair if it contains men and women in equal proportion. We adopt the same preprocessing as \cite{xu2020algorithmic}. After preprocessing, this dataset contains $39190$ items and $112$ features. We set sex as the protected attribute. We use this dataset with talent searching in mind.
    \item \textbf{MovieLens100k} \cite{harper2016movielens}. In this dataset, an item represents a movie. It contains $943$ users, $1682$ items, and $100000$ interactions in total. Each user has at least $20$ interactions. We consider two protected attributes. The first is based on periods of movies. Some recommender systems may recommend only new movies, but some users may want to know old movies as well. We divide movies into two groups, such that the protected group contains movies that were released before 1990. The second protected attribute is based on popularity. Movies that received less than $50$ interactions are considered to be in the protected group. 
    \item \textbf{LastFM} \footnote{\url{https://grouplens.org/datasets/hetrec-2011/}}. In this dataset, an item represents a piece of music. To discard noisy items and users, we extract $10$-cores of the dataset, such that each user and item have at least $10$ interactions. Specifically, we iteratively discard items and users with less than $10$ interactions until all items and users have at least $10$ interactions. It contains $1797$ users, $1507$ items, and $62376$ interactions in total after preprocessing. We consider popularity as a protected attribute. The protected group contains music that received less than $50$ interactions.
    \item \textbf{Amazon Home and Kitchen} \cite{he2016ups, mcauley2015image}. In this dataset, an item represents a home and kitchen product on amazon.com. We extract $10$-cores of the dataset using the same preprocessing as in the LastFM dataset. It contains $1395$ users, $1171$ items, and $25445$ interactions in total after preprocessing. We consider popularity as a protected attribute. The protected group contains products that received less than $50$ interactions.
\end{itemize}

\end{document}